\documentclass[pra,twocolumn,superscriptaddress,10pt]{revtex4-2}
\usepackage{amssymb,amsmath,color,mciteplus,graphicx,subfigure}
\usepackage{calc,epsfig,epstopdf,color,mciteplus,bm,mathrsfs}
\usepackage{times}
\usepackage{float}
\usepackage{multirow}
\usepackage{lipsum}
\usepackage{ulem}

\usepackage[ruled,vlined,linesnumbered]{algorithm2e}

\begin{document}
%\preprint{APS/123-QED}
\title{Exploring critical states of the quantum Rabi model via Hamiltonian variational ansätze}
\author{Mei Peng}
\affiliation{Key Laboratory of Atomic and Subatomic Structure and Quantum Control (Ministry of Education),  Guangdong Basic Research Center of Excellence for Structure and Fundamental Interactions of Matter, and  School of Physics, South China Normal University, Guangzhou 510006, China}

\author{Xu-Dan Xie}
\affiliation{Key Laboratory of Atomic and Subatomic Structure and Quantum Control (Ministry of Education),  Guangdong Basic Research Center of Excellence for Structure and Fundamental Interactions of Matter, and  School of Physics, South China Normal University, Guangzhou 510006, China}

\author{Dan-Bo Zhang} \email{dbzhang@m.scnu.edu.cn}
\affiliation{Key Laboratory of Atomic and Subatomic Structure and Quantum Control (Ministry of Education),  Guangdong Basic Research Center of Excellence for Structure and Fundamental Interactions of Matter, and  School of Physics, South China Normal University, Guangzhou 510006, China}
\affiliation{Guangdong Provincial Key Laboratory of Quantum Engineering and Quantum Materials,  Guangdong-Hong Kong Joint Laboratory of Quantum Matter, and Frontier Research Institute for Physics,\\  South China Normal University, Guangzhou 510006, China}
\begin{abstract}
Characterizing quantum critical states towards the thermodynamic limit is essential for understanding phases of matter. The power of quantum simulators for preparing the critical states relies crucially on the structure of quantum circuits and in return provides new insight into the critical states. Here, we explore the critical states of the quantum Rabi model~(QRM) by preparing them variationally with Hamiltonian variational ansätze~(HVA), in which the intricated interplay among different quantum fluctuations can be parameterized at different levels.  We find that the required circuit depth scales linearly with the effective system size, suggesting that HVA can efficiently capture the behavior of critical states of QRM towards the thermodynamic limit. Moreover, we reveal that HVA gradually squeeze the initial state to the target critical state, with a number of blocks increasing only linearly with the effective system size. Our work suggests variational quantum algorithm as a new probe for the complicated critical states. 
\end{abstract}
\maketitle

\section{Introduction}\label{sec:level1}
Quantum phase transitions (QPT) describe abrupt changes in the ground-state properties of quantum many-body systems driven by quantum fluctuations~\cite{vojta2003quantum,heyl2018dynamical}, which play a crucial role in understanding the critical behavior of strongly correlated systems~\cite{bakemeier2012quantum,puebla2013excited,hwang2016quantum,ashhab2010qubit,ashhab2013superradiance,puebla2017probing,hwang2015quantum,puebla2016excited}. The critical state describing the transition point shows scaling behavior towards the thermodynamic limit, typically in terms of the system size. Yet there are other types of systems with no system sizes; the thermodynamic limit can still be defined. For instance, the quantum Rabi model (QRM), which consists of a single-mode cavity field coupled to a two-level atom, also exhibits quantum phase transitions~\cite{hwang2015quantum,puebla2016excited,chen2024sudden,chen2024error}, with a thermodynamic limit defined as the ratio of the atomic transition frequency $\Omega$ and the cavity field frequency $\omega_0$ approaching infinity.
Characterizing and efficiently preparing critical states of QRM not only enrich our understanding of novel physics in light-atom systems,  but can also provide resources for quantum metrology~\cite{liu2022process,gietka2023unique}. 

Rapid advances in quantum simulators allow us to investigate the properties of critical states by preparing them directly~\cite{feynman2018simulating,lloyd1996universal,greiner2002quantum,blatt2012quantum,ebadi2021quantum,bakemeier2012quantum}.  The variational quantum eigensolver (VQE) is widely regarded as one of the few efficient algorithms currently suitable for near-term quantum simulators~\cite{cerezo2021variational,preskill2018quantum,bharti2022noisy}. It can achieve high accuracy with fewer qubits and shallower circuits, by leveraging hybrid quantum-classical optimization to steer a trial state toward the target state~\cite{farhi2014quantum,farhi2016quantum,peruzzo2014variational,nakanishi2019subspace,wecker2015progress,kokail2019self}.  The efficiency of the algorithm strongly depends on the choice of ansätze, which often need to be tailored to the specific problem. The multi-scale entanglement renormalization ansätze is well-known  for capturing the entanglement structure of critical states of quantum lattice systems~\cite{evenbly2013quantum,arguello2022generalized,luchnikov2021simulating}. Alternatively, it is also shown that the Hamiltonian variational ansätze~(HVA) can prepare critical states of the transverse-field Ising model~(TFIM) , with a circuit depth only half of the system size~\cite{ho2019efficient}. However, 
there is less progress for preparing critical states of the QRM with digital quantum simulation. One major difficulty lies in simulating hybrid-variable quantum systems. Since the cavity mode is a continuous variable, mapping it onto qubits often demands a large number of qubits and leads to complex quantum circuits.  The challenge can be overcome by directly using a hybrid-variable quantum simulator~\cite{peruzzo2014variational,nakanishi2019subspace,wecker2015progress,kokail2019self,ho2019efficient,lloyd2003hybrid,lloyd1999quantum,zhang2021continuous,wang2025quantum,dutta2025solving,araz2024toward}. Nevertheless, even with hybrid-variable quantum simulators at hand, it remains unclear whether the critical state of the QRM can be efficiently prepared in the thermodynamic limit. If the answer is yes, can the ansätze itself reveal some intrinsic structure of the critical states? 

In this work, we investigate the critical states of the QRM using a hybrid-variable variational quantum algorithm, aiming to efficiently prepare the critical state with an efficient ansätze and simultaneously reveal its underlying structure.  By decomposing the Hamiltonian as different components, we can adopt the HVA with multiple blocks.  Each block captures the competition between quantum fluctuations of different components by alternately evolving each part of the Hamiltonian with parameterized time durations. We find that the variational quantum algorithm with HVA can efficiently prepare the critical states of the QRM with high accuracy in the thermodynamic limit. Notably, the required number of ansätze blocks increases only linearly with the effective system size~(logarithm of the ratio $\frac{\Omega}{\omega_0}$). In this regard, HVA fine-tunes the competition among quantum fluctuations block by block until the target critical state can be obtained. Remarkably, we observe that each block of HVA effectively squeezes the state with almost the same squeezing factor. This can explain the linear scaling of block number towards the thermodynamic limit as those blocks can produce the required squeezing for the critical state. Our work suggests the HVA as an efficient description as well as a useful tool for investigating the critical states on near-term quantum simulators.  

The structure of this paper is as follows.  Sec.~\ref{sec:level2} provides a brief introduction to the Hamiltonian of the QRM and presents a variational quantum algorithm suitable for the QRM.
In Sec.~\ref{sec:level3}, we analyze the critical state via the variational quantum algorithms by numerical simulations. Finally, Sec.~\ref{sec:level4} summarizes the conclusions.

\section{Hamiltonian variational ansätze for the quantum Rabi model}\label{sec:level2}
In this section, we  provide a brief introduction to the QRM and propose the HVA suitable for solving its critical state.  

\subsection{Quantum Rabi model}
The QRM describes the interaction between a two-level system (such as a qubit) and a single-mode quantum resonant cavity field. Its Hamiltonian is given by (with $\hbar=1$):
\begin{eqnarray}\label{eq:ham}
H_{Rabi}=\omega_{0}a^{\dagger}a+\frac{\Omega}{2}\sigma_{z}-\lambda(a^{\dagger}+a)\sigma_{x},
\end{eqnarray}
where $\omega_0$ is the cavity mode frequency, $\Omega$ is the atomic transition frequency, and $\lambda$ is the coupling strength. The QRM has a  $Z_2$ symmetry, defined as $\Pi = e^{i\pi a^\dagger a} \sigma_z$. When $\lambda = 0$, the eigenstate of the QRM is the direct product state of the photon number state $\vert{n}\rangle$ and the spin state $\vert{\sigma}\rangle$, expressed as $\vert{n,\sigma\rangle}$. Conversely, for $\lambda \neq 0$, the eigenstates are typically entangled superpositions of these direct product states. This entanglement signifies a strong interaction between the photonic field and the spin state, thereby precluding a simple, separable description and rendering the calculation and simulation of the QRM considerably challenging.

The QRM has no system size and thus the conventional thermodynamic limit fails to work here. However, the ratio $\frac{\Omega}{\omega_0}$ approaching infinity can effectively act as a thermodynamic limit. In the large $\frac{\Omega}{\omega_0}$ limit, the Schrieffer-Wolff transformation $U=e^{\frac{\lambda}{\Omega}(a+a^{\dagger})(\sigma_+-\sigma_-)}$ can be employed to solve the Hamiltonian efficiently~\cite{braak2011integrability,chen2024sudden}. By transforming $U^{\dagger}H_{Rabi}U$, the coupling term between the spin subspaces $H_\downarrow$ and $H_\uparrow$ in the Hamiltonian becomes negligible. When projecting onto $H_\downarrow$, i.e., defining
$H_{eff}=\langle{\downarrow}|U^{\dagger}H_{Rabi}U|\downarrow\rangle$, an effective low-energy Hamiltonian is obtained as, 
\begin{equation}\label{eq:H_eff}
H_{eff}=\omega_0a^{\dagger}a-\frac{\omega_0g^2}{4}(a+a^{\dagger})^2-\frac{\Omega}{2},
\end{equation}
where $g=\frac{2\lambda}{\sqrt{\omega_0\Omega}}$. When $g=1$, a QPT occurs accordingly. At this point, the low-energy eigenstates are squeezed states $\vert\varphi\rangle=S(x)\vert{n}\rangle\vert{\downarrow}\rangle$ with the squeezing operator $S(x)=e^{\frac{x}{2}(a^{\dagger}{^2}-a^2)}$ and $x=-\frac{1}{4}\ln{(1-g^2)}$\cite{hwang2015quantum,puebla2016excited}. The $Z_2$ symmetry is broken for $g>1$.

Typically, simulating quantum states at critical points is relatively difficult. Moreover, it is crucial to note that quantum computers are mostly based on qubits. Therefore, to simulate quantum systems with continuous variables,  the creation operator $a^{\dagger}$ and the annihilation operator $a$ in the Hamiltonian must be represented in terms of Pauli operators. However, this mapping typically requires a large number of qubits and complex quantum circuits. An alternative approach is to directly encode continuous variables into physical systems that inherently support them~\cite{andersen2015hybrid}. For simulating the QRM, a more natural solution is to adopt a hybrid quantum simulator, in which the two-level atomic system is represented by a qubit, while the cavity mode is encoded as a continuous variable. Such a hybrid quantum simulator exists naturally in quantum platforms such as superconducting systems~\cite{wallraff2004strong,paik2011observation,devoret2013superconducting} and trapped ions~\cite{sutherland2021universal,gan2020hybrid,monroe2013scaling}. 

\begin{figure}[htbp]
 \centering
 \includegraphics[scale=0.24]{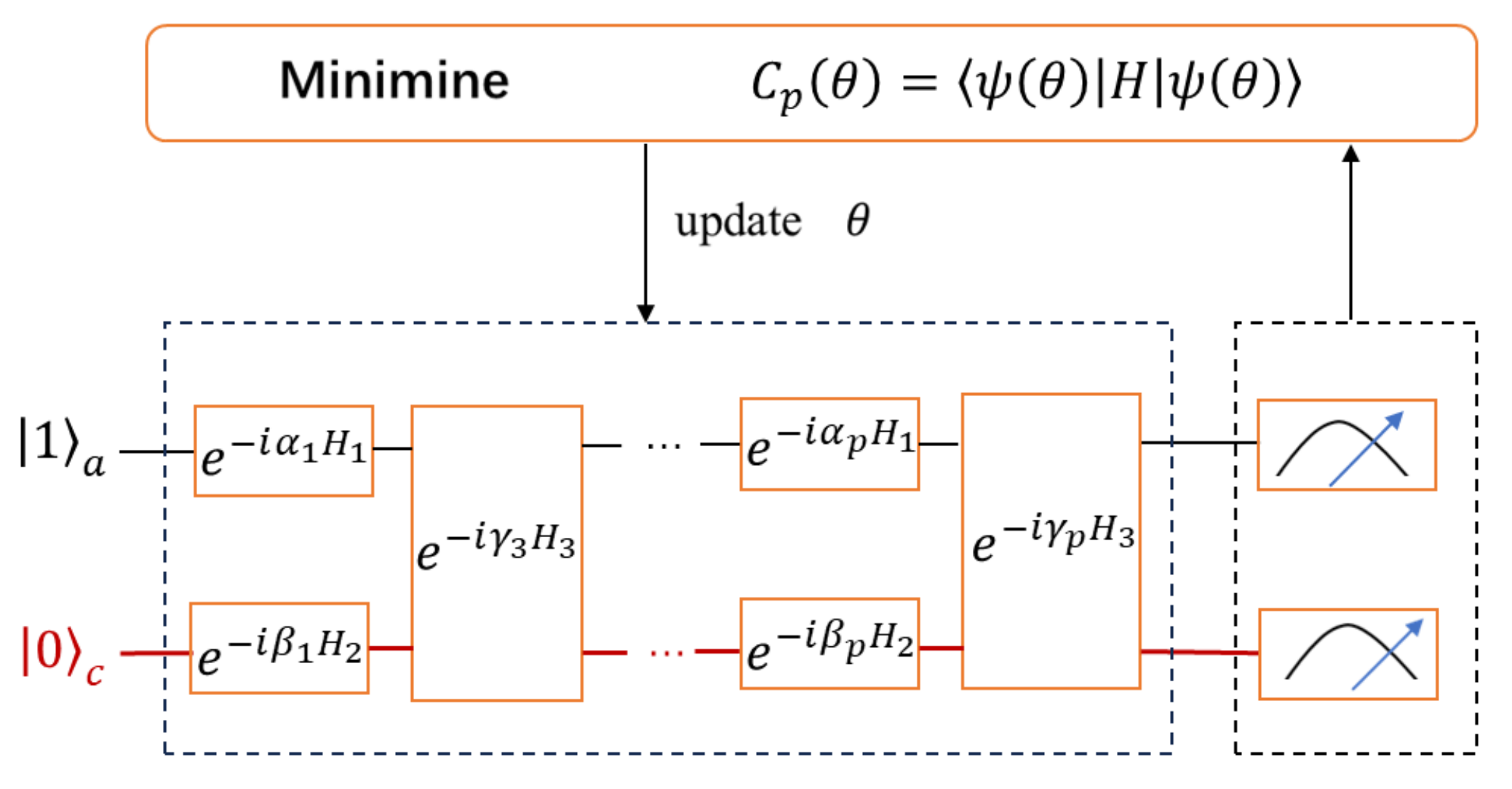}
\caption{ \textbf{Illustration of VQE with hybrid-variables.} The black and red lines represent the qubit and the continuous-variable, respectively. The HVA consists of blocks of parameterized evolutions of $H_1$, $H_2$,$H_3$, a decomposition of the Hamiltonian of the QRM.}\label{Fig:qc} 
\end{figure}

\subsection{variational quantum eigensolver}
Critical states at QPT points are typically difficult to obtain due to their intricate structure. The VQE has emerged as a promising approach for preparing ground states of quantum many-body systems.  Remarkably, using the HVA, it is shown that VQE can efficiently prepare critical states of the TFIM~\cite{ho2019efficient}. 

To prepare the critical states of the QRM, we extend the application of the VQE to hybrid-variable quantum computers. This extension is straightforward by introducing variational parameters into both continuous-variable and hybrid-variable quantum gates. While a general framework may be established, here we restrict our discussion to the QRM. To obtain a good approximation of the ground state and to alleviate the impact of barren plateaus during optimization, we express the target many-body Hamiltonian as a linear combination of $H_1$, $H_2$, and $H_3$, and employ the HVA to construct the quantum circuit~\cite{wiersema2020exploring}. The HVA for the QRM is as follows:
\begin{eqnarray}
&&U(\theta)=\prod\limits_{j=1}^{p}U(\theta_j),\nonumber \\
&&U(\theta_j)=e^{-i\gamma_jH_3}e^{-i\beta_jH_2}e^{-i\alpha_jH_1},
\label{eq:QCM}  
\end{eqnarray}
where $H_1=\sigma_z$, $H_2=a^{\dagger}a$, $H_3=(a+a^{\dagger})\sigma_x$, $\theta_j=(\alpha_j,\beta_j,\gamma_j)$ are the set of variational parameters. Note that $e^{-i\gamma_jH_3}$ is the hybrid-variable gate, which couples the qubit and the continuous-variable with a tunable parameter $\gamma_j$. 

As illustrated in Fig.~\ref{Fig:qc}, the VQE employs a parameterized quantum circuit to implement the unitary transformation $U(\theta)$, which generates a trial state $\vert{\psi(\theta)\rangle=U(\theta)}\vert{\psi_0}\rangle$ as a candidate critical state. The initial state $\vert \psi_0 \rangle$ is chosen as a product of the vacuum state of the bosonic mode and the spin-down state of the qubit, i.e., $\vert \psi_0 \rangle = \vert 0 \rangle_c \otimes \vert 1 \rangle_a$. The cost function, defined as the expected value of the Hamiltonian of the QRM,  $C_p(\theta)=\langle\psi(\theta)\vert{H}\vert\psi(\theta)
\rangle$, is then evaluated. A classical optimizer updates the parameters to generate a new set $\theta_p$, which is fed back into the quantum circuit to generate a new trial state. This iterative process continues until $C_p(\theta)$ reaches its minimum. At this moment, the state corresponding to the optimized $\theta_p$ approximates the ground state of the target Hamiltonian. Next, we evaluate the fidelity and analyze the circuit resources required to prepare the critical ground state using the algorithm.

\section{Analysis of critical states with Hamiltonian variational ansätze}\label{sec:level3}
We now employ VQE with the HVA to investigate the critical states of the QRM for different ratios of $\Omega/\omega_0$.  We first present how fidelities between the critical states obtained by VQE and the exact ones increase with the depth of quantum circuits. Then we explore the nature of the critical state by analyzing the photon number distribution and analyze its characteristics from the perspective of effective squeezing factors.

In our numerical analysis, we fix $\omega_0=0.1$. Since the ratio  $\frac{\Omega}{\omega_0}$ effectively determines the dimensionality of the system's Hilbert space (approximately scaling as $\frac{\Omega}{\omega_0}\sim2^L$), we emulate the thermodynamic limit condition ($\frac{\Omega}{\omega_0}\to \infty$) by allowing $\Omega$ to grow exponentially, e.g., $\Omega=2^{i}(i=2,3,4\cdots)$. The condition $g = 1$ as the critical point is preserved for different $\frac{\Omega}{\omega_0}$  by setting proper $\lambda$ accordingly.  The numerical simulation is performed using the open-source package \textit{QuTiP}~\cite{johansson2012qutip}.

\subsection{Fidelity}
We first evaluate the fidelity $F_p(\theta)=|\langle\psi_t|\psi_p\rangle|^2$ between the variationally prepared state $|\psi_p\rangle$ and the exact ground state $|\psi_t\rangle$  of the QRM Hamiltonian, across different values of the ratio $\frac{\Omega}{\omega_0}$. Since the Fock space is theoretically infinite, truncation is necessary for practical calculations. In this study, a photon number truncation of $N=60$ is applied, which has been checked to be accurate enough for a critical state at $\Omega=64$, which is the largest value used in our simulation. 

\begin{figure}[htbp]
 \centering
 \includegraphics[scale=0.42]{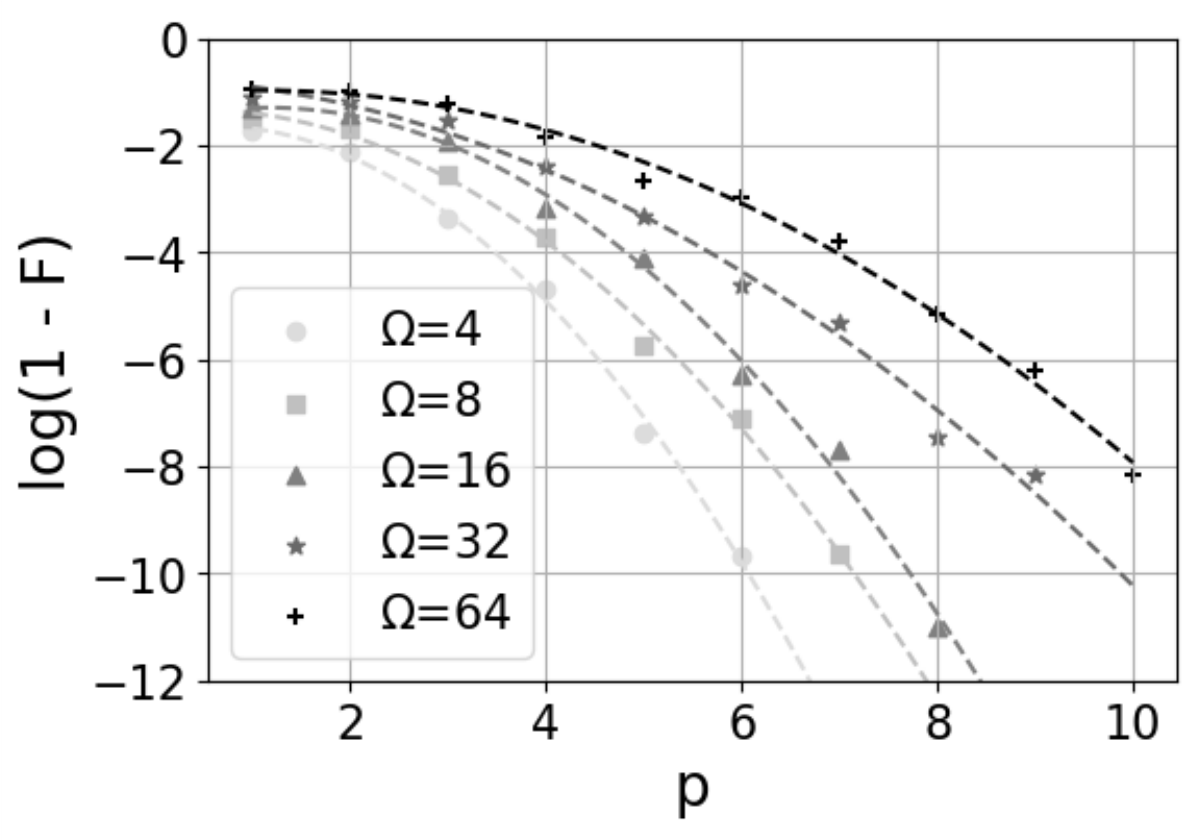}
\caption{Relationship between the logarithmic infidelity $\log(1-F)$ and the number of circuit layers $p$, under various $\Omega$. The parameters are set at the critical point $g=1$, with $\omega_0=0.1$, $\Omega=2^{i}$ ($i=2,3,4\cdots$) , and $\lambda=\frac{g\sqrt{\omega_0\Omega}}{2}$. }\label{Fig:fidelity}
\end{figure}

\begin{figure}[htbp]
 \centering \includegraphics[scale=0.55]{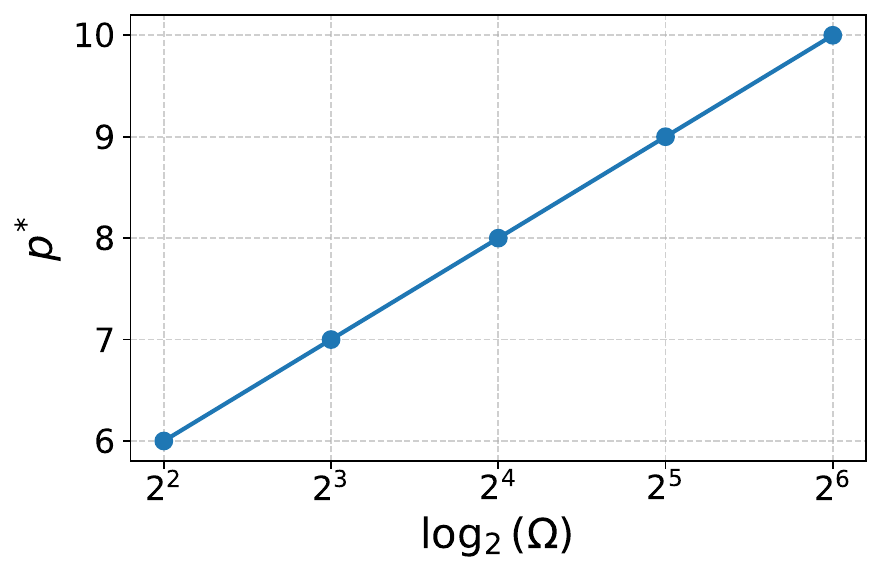}
\caption{The linear scaling relation between the required circuit depth $p^*$ for accurately preparing the critical state and the effective system size $\log_2(\Omega)$. }
\label{Fig:scalling}
\end{figure}

Fig.~\ref{Fig:fidelity} uncovers the relationship between the fidelity and the circuit depth. The results show that that the fidelity improves progressively as the number of circuit layers increases for all values of  $\Omega$. This indicates that, as the complexity and expressiveness of the circuits grow, the critical state can be obtained by the HVA with greater accuracy.  In addition, as $\Omega$ grows exponentially, the required circuit depth for accurately preparing the critical state also increases, but only moderately. By setting a threshold of infidelity $10^{-7}$ and denoting the required circuit depth as $p^*$, as illustrated in the fig.~\ref{Fig:scalling}, the required circuit depth scales linearly with $\log(\Omega)$, i.e., $p^*\sim \log_2\Omega$. If we interpret the ratio $\frac{\Omega}{\omega_0}$ as an effective dimension of Hilbert space, then $\log_2\frac{\Omega}{\omega_0}$ is an effective system size. The required circuit depth thus is dependent only linearly with the effective system size. Such a scaling behavior shows that the HVA can efficiently express the critical state of the QRM, demonstrating its power of simulating critical states beyond the TFIM.  

\begin{figure*}[htbp]
 \centering
 \includegraphics[scale=0.45]{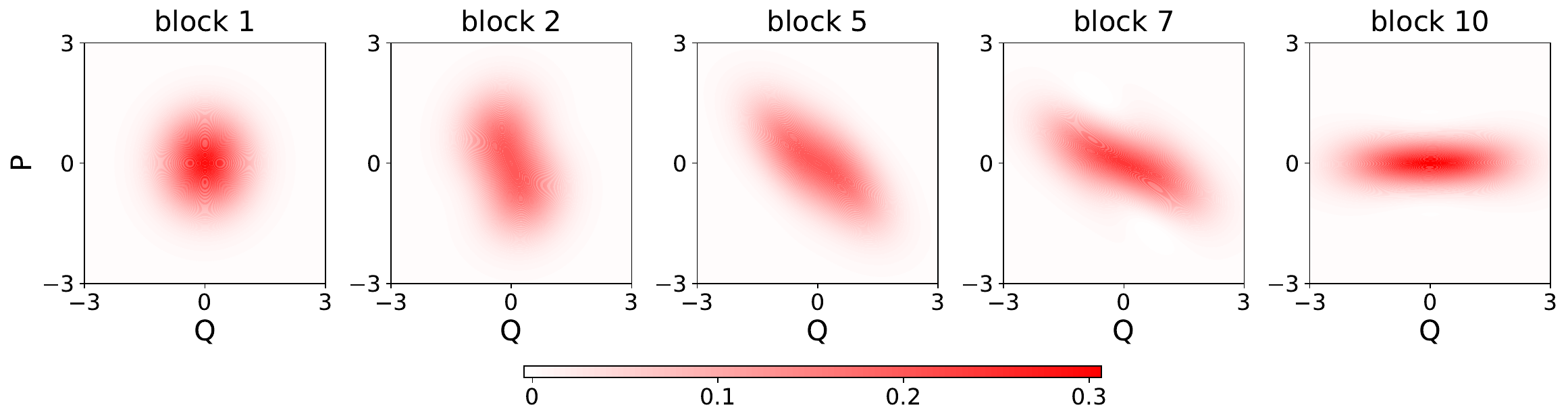}
 \caption{ \textbf{Wigner probability distribution.} The process of state generation with successive blocks ($j=1,2,5,7,10$) along the quantum circuit. The horizontal and vertical axes represent the position $Q$ and momentum $P$, respectively. The initially symmetric, disk-like Wigner distribution is progressively squeezed and rotated toward the target state as the circuit depth increases. Here, the parameters are $\Omega=64$, $\omega_0=0.1$ and $\lambda=1.26$. The total number of blocks is $10$.}
 \label{Fig:Wigner}
\end{figure*}

%The numerical fit of Fig.~\ref{Fig:fidelity} further illustrates in detail the quadratic relationship between fidelity and the number of circuit layers. In this model, the logarithm of the fidelity is related to the square of the number of circuit layers, specifically: $\log(1-F)=ap^{2}+bp+c$. This quadratic relationship indicates that the improvement in fidelity accelerates as the circuit depth increases. Additionally, this formula shows that, as the number of circuit layers increases, the quantum algorithm can more accurately approach the true eigenstates of the quantum system, ultimately bringing the fidelity closer to 1 and enabling the precise preparation of the quantum state. This quadratic relationship holds significant implications for quantum computing, as it not only establishes the connection between quantum circuit depth and fidelity, but also reveals that as the system size grows, the efficiency of variational quantum algorithms will increase at an accelerating rate.

%In summary, the analysis in Fig.~\ref{Fig:fidelity} not only shows the direct relationship between quantum circuit depth and system fidelity, but also reveals the logarithmic growth relationship between circuit depth and system parameters as the system scale increases, providing an important theoretical basis for the application of variational quantum algorithms in processing large-scale quantum systems. These results further demonstrate the close connection between the complexity of quantum circuits and the expressive power of the system.

\subsection{Generating the critical state with successive blocks}
We now analyze how the quantum circuit employing the HVA with optimized parameters generates the critical state. We focus on the case of $\Omega = 64$, the largest one investigated in this work, and consider $p = 10$ blocks, which can reach almost the perfect fidelity. 
To elucidate the state generation process, the intermediate wavefunctions along the quantum circuit are extracted and analyzed. It is more informative to focus on the continuous-variable component of the state.
We first present the Wigner probability distribution corresponding to the continuous-variable density matrix generated after the first $j$ blocks. For a pure state with wavefunction $\psi(Q)$, the Wigner probability distribution can be obtained by the Wigner function $W(Q,P)$, which is defined as~\cite{hillery1984distribution} :
\begin{equation}
W(Q,P) = \frac{1}{\pi \hbar} \int_{-\infty}^{\infty} \psi^*\left(Q + \frac{q}{2}\right) \psi\left(Q - \frac{q}{2}\right) e^{i P q / \hbar} \, dq.
\end{equation}
The Wigner probability distribution illustrates the phase-space structure of a continuous-variable quantum state, with $Q$ and $P$ denoting the position and the momentum, respectively. 
As shown in Fig.~\ref{Fig:Wigner}, as the circuit depth increases, the initially symmetric Gaussian distribution becomes progressively squeezed and rotated. By the 10th block, strong single-axis squeezing emerges, indicating that the quantum state evolves toward a critically squeezed state. 

\begin{figure}[hbp]
 \centering
\includegraphics[scale=0.42]{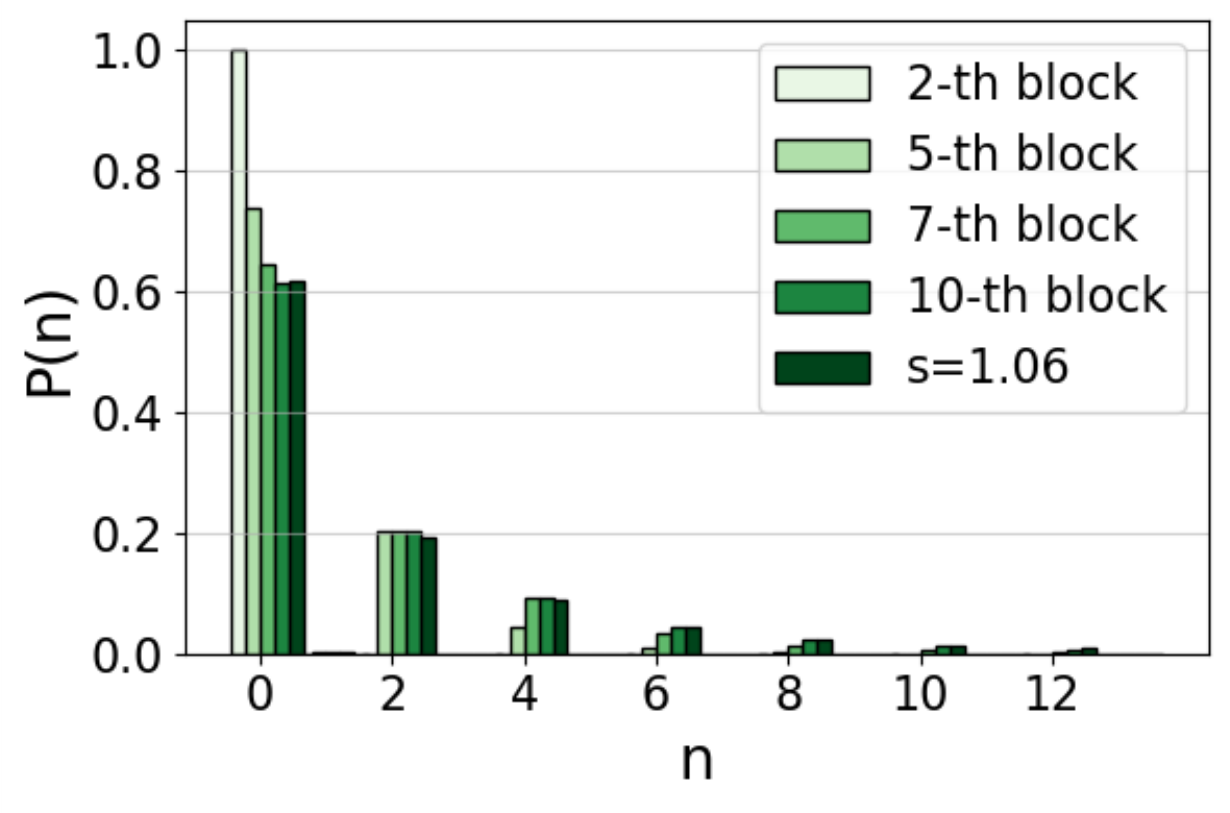}
 \caption{ \textbf{Fock number distribution for quantum states generated with successive blocks in the quantum circuit. }
The parameters are $\Omega=64$, $\omega_0=0.1$ and $\lambda=1.26$. The total number of blocks is $10$.}
 \label{Fig:pn}
\end{figure}

We also provide another aspect for the generation of the critical state along the successive blocks of HVA via the Fock number distribution. 
As shown in Fig.~\ref{Fig:pn}, the system initially resides almost entirely in the vacuum state. As the circuit depth increases, the probability of the vacuum state gradually decreases, and the population shifts to higher Fock states.  Remarkably, it is interesting to see that the distribution consists of only even numbers. This behavior arises from the fact that the initial state has even parity and the parameterized quantum circuit preserves the $Z_2$ symmetry inherited from the symmetry of each decomposed term in the Hamiltonian. In this regard, the HVA is symmetry-preserving.

\begin{figure*}[htp]
 \begin{minipage}{0.32\linewidth}
 	\centerline{\includegraphics[width=\textwidth]{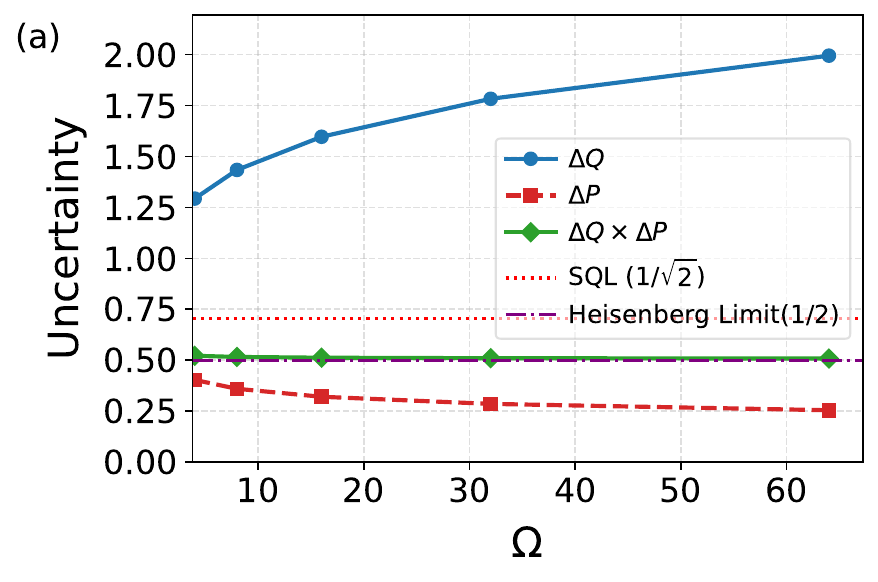}}	
 \end{minipage}
 \begin{minipage}{0.32\linewidth}
  \centerline{\includegraphics[width=\textwidth]{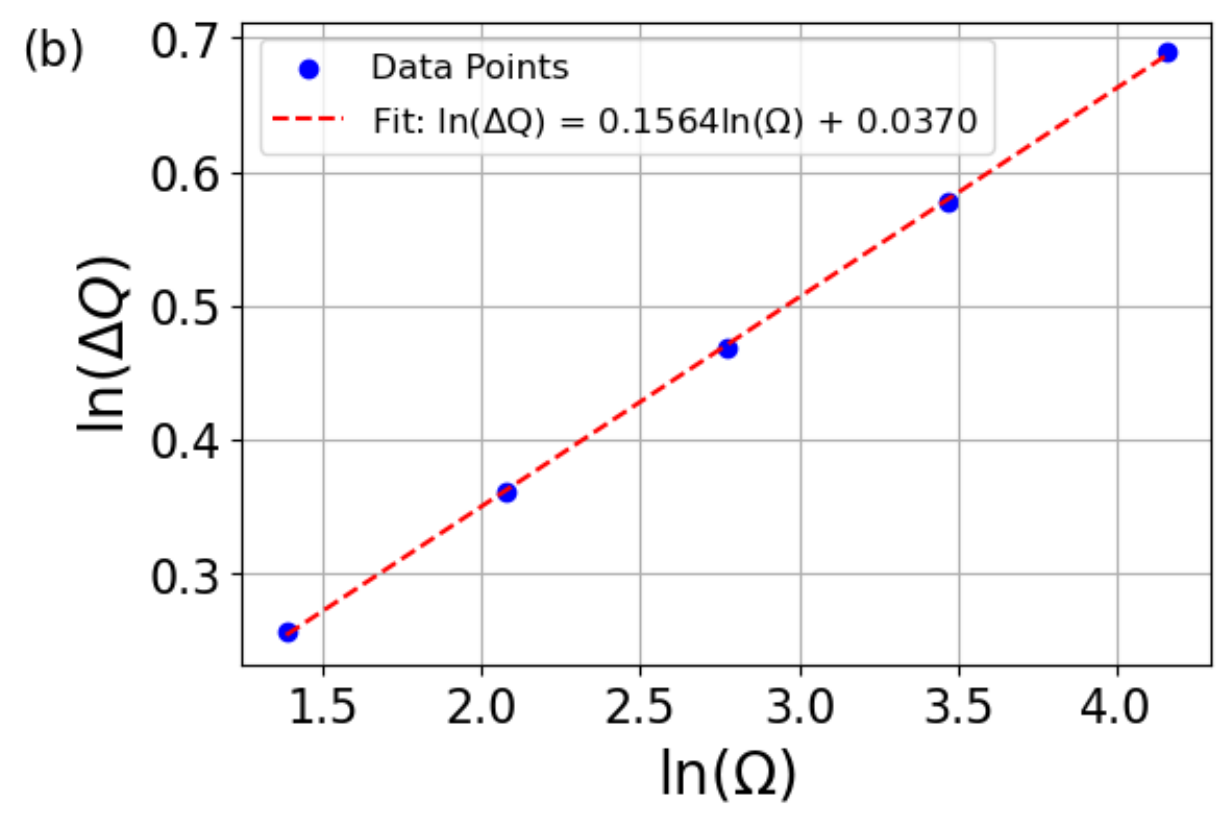}}
 \end{minipage}
 \begin{minipage}{0.32\linewidth}    \centerline{\includegraphics[width=\textwidth]{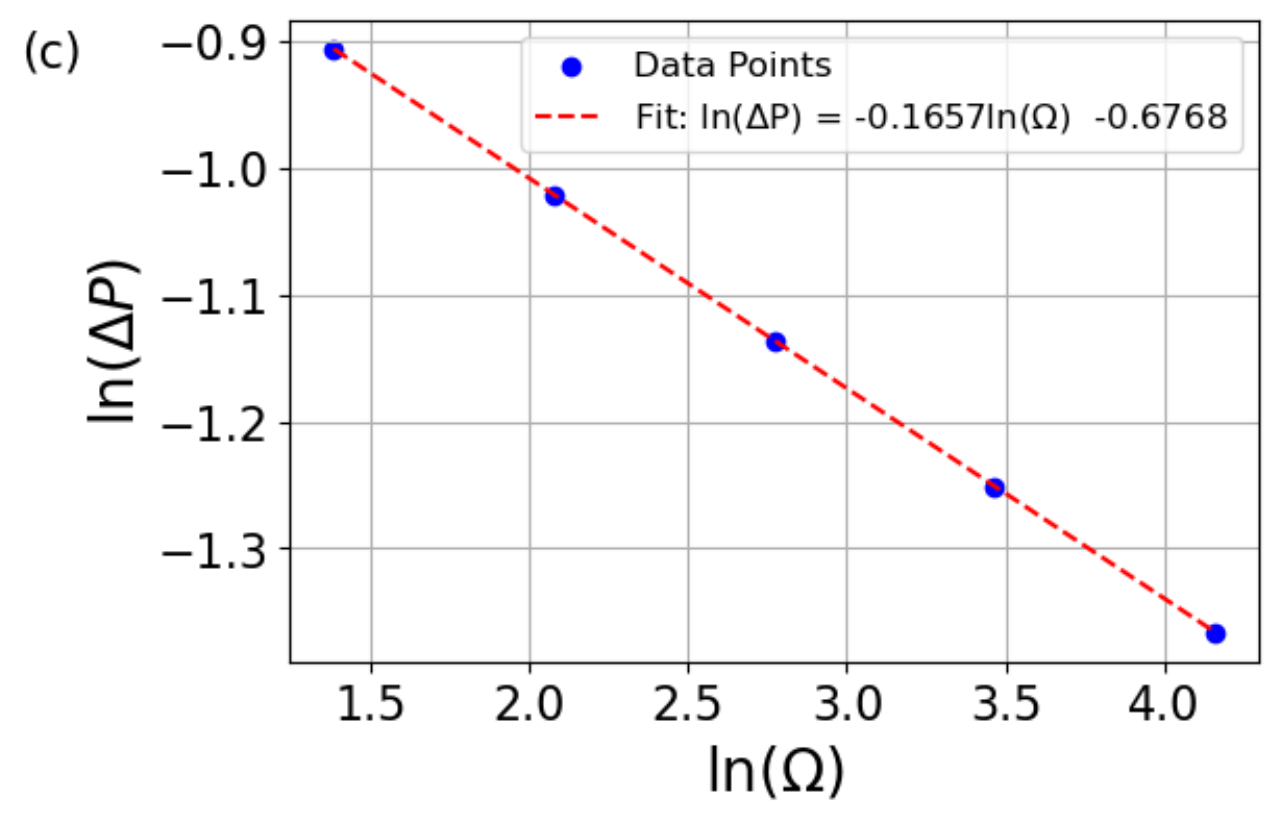}}	
 \end{minipage}
	\caption{\textbf{Precision solutions for $\Delta{Q}$ and $\Delta{P}$.} (a) Fluctuations in position and momentum directions under different $\Omega$, and the variation in uncertainty. (b) The relationship between the $\ln{\Delta{Q}}$ and $\ln{\Omega}$. (c) The relationship between the $\ln{\Delta{P}}$ and $\ln{\Omega}$. The parameter is $\Omega=2^{i}$, where $i=2,3,4\cdots$.}\label{Fig:pre}
\end{figure*}

\begin{figure*}[htp]
 \begin{minipage}{0.48\linewidth}
 	\centerline{\includegraphics[width=\textwidth]{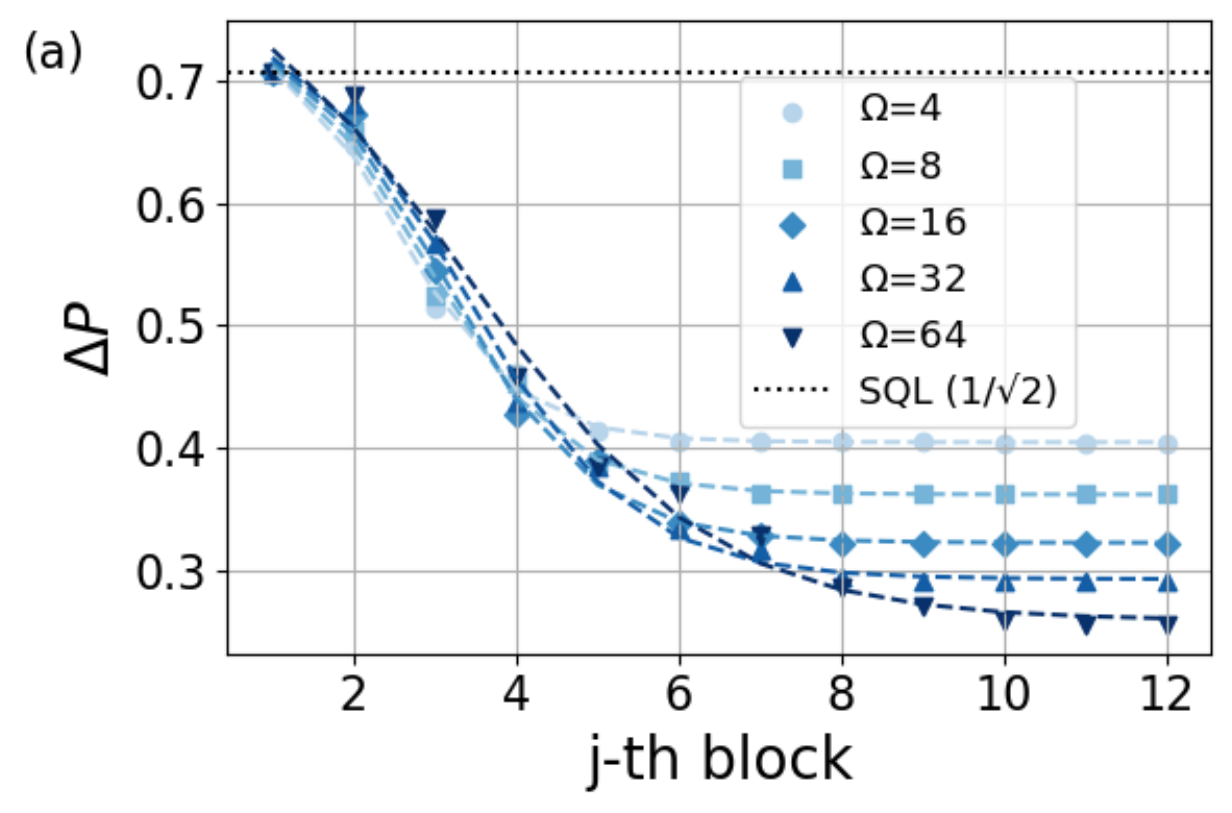}}    
 \end{minipage}
 \begin{minipage}{0.48\linewidth}    \centerline{\includegraphics[width=\textwidth]{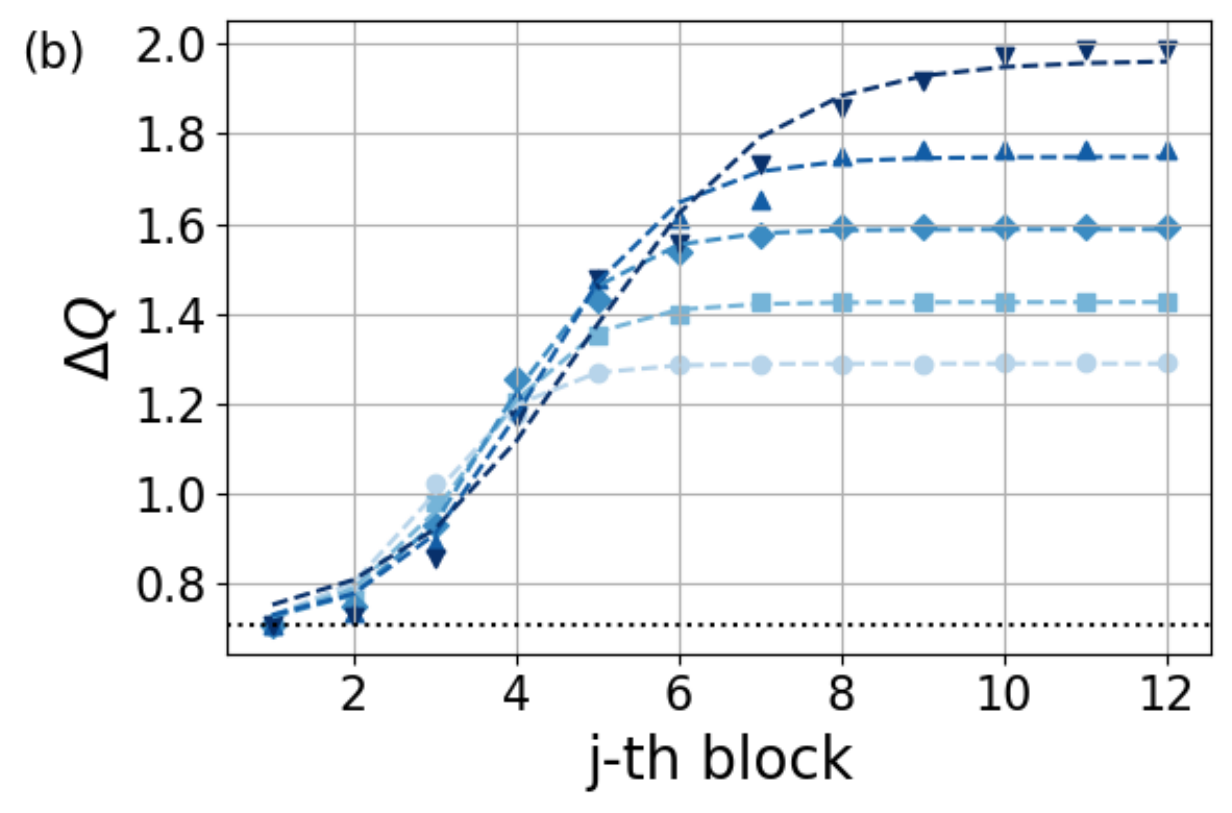}}    
 \end{minipage}
 
	\caption{\textbf{VQA solutions for $\Delta{Q}$ and $\Delta{P}$.} (a) Standard deviation of momentum, $\Delta P$, as a function of variational circuit depth for different values of $\Omega$. In all cases, $\Delta P$ decreases with increasing depth and eventually saturates below the standard quantum limit (SQL $= 1/\sqrt{2}$), demonstrating strong quantum squeezing and the block-by-block squeezing mechanism of the HVA.
    (b) Standard deviation of position, $\Delta Q$, as a function of variational circuit depth. For all values of $\Omega$, $\Delta Q$ increases with depth and saturates above the SQL, indicating significant quantum anti-squeezing. }\label{Fig:SPp}
    % $\Delta{Q}$ and $\Delta{P}$ follows a almost linear relation with the number of blocks until convergence.
\end{figure*}

As the circuit depth increases, the system exhibits transitions between even Fock numbers, a phenomenon that is identical to the photon number distribution in the squeezed vacuum state, where only even numbers can be detected \cite{breitenbach1997measurement}. As shown in Fig.~\ref{Fig:pn}, we can confirm that the critical state of the QRM is the squeezing vacuum state. The process can only generate a squeezed vacuum state if the Hamiltonian of the system describes a two-photon process that evolves over time. As introduced in Sec.~\ref{sec:level2} for the QRM, while $H_{Rabi}$ represents a single-photon process, it achieves the outcome of a two-photon process in $H_{eff}$ as defined in Eq.~\eqref{eq:H_eff}. 

From the previous analysis, we know that the quantum circuit we designed did not directly introduce the squeezing operator initially, yet it was still able to successfully prepare the squeezed state ultimately. This suggests that, although the squeezing operator was not explicitly applied, the quantum gates and circuit operations in the design effectively played a role similar to the squeezing operator. Specifically, the optimization process at each layer corresponds to applying a squeeze to the system, gradually transforming the quantum state from the initial vacuum state to the final critical state. Through the block-by-block optimization of the quantum circuit, the system undergoes successive squeezing, ultimately resulting in the final critical state exhibiting the characteristics of a squeezed state, where even photon numbers dominate and odd photon numbers nearly vanish.

%This discovery is highly significant, as it demonstrates that by properly designing quantum circuits, we can prepare squeezed states without directly introducing squeezing operators. The optimization of quantum gates at each layer not only enhances the system's fidelity but also effectively squeezes the system's state to the desired critical state, ultimately achieving the squeezing characteristics of the QRM critical state. The successful implementation of this process validates the powerful utility of quantum gates in circuit design and highlights the potential of quantum circuits for quantum state preparation.

\subsection{Analysis of squeezing factor}
To further quantify the squeezing behavior observed in the evolution of phase-space structures, we examine the variation of the squeezing factor across successive blocks in the HVA.

We first introduce the canonical variable operators—the position operator $\hat{Q}$ and the momentum operator $\hat{P}$, defined respectively as,
\begin{eqnarray}
&\hat{Q}=\frac{1}{\sqrt{2}}(a+a^{\dagger}), \nonumber\\
&\hat{P}=-\frac{i}{\sqrt{2}}(a-a^{\dagger}).
\label{eq:PQ}
\end{eqnarray}
According to the fundamental principles of quantum mechanics, the canonical variable operators $\hat{Q}$ and $\hat{P}$ satisfy the commutation relation $[\hat{Q},\hat{P}]=i$, meaning that they cannot be measured simultaneously with perfect precision. This uncertainty relation is particularly pronounced in squeezed states: when fluctuations in one variable are suppressed, fluctuations in the other are necessarily amplified. To quantitatively characterize the squeezing properties of the QRM critical state, we define the standard deviations of the position and momentum operators,
\begin{eqnarray}
&\Delta{Q}=\sqrt{\langle{\hat{Q}^2}\rangle_\rho-\langle{\hat{Q}}\rangle{^2}_\rho},\nonumber\\
&\Delta{P}=\sqrt{\langle{\hat{P}^2}\rangle_\rho-\langle{\hat{P}}\rangle{^2}_\rho}.
\label{eq:VPQ}
\end{eqnarray}
where $\rho=|\psi_p\rangle\langle\psi_p|$ is the system's density operator. For an ideal squeezed state, the uncertainty relation   $\Delta{Q}\Delta{P}$  is equal to $\frac{1}{2}$. When $\Delta{Q}$ or $\Delta{P}$ is smaller than $\frac{1}{\sqrt{2}}$, then $\hat{Q}$ or $\hat{P}$ is squeezed.  

Numerical results reveal that the quantum fluctuations of the QRM critical state exhibit significant anisotropy: a pronounced squeezing effect is observed in the momentum direction, accompanied by amplified fluctuations in the position direction. Notably, as $\Omega$ increases, the degree of squeezing and amplification intensifies, yet the system consistently maintains an approximately constant uncertainty relation $\Delta{Q}\Delta{P}\approx\frac{1}{2}$. Further analysis (Fig.~\ref{Fig:pre}(b) and (c)) shows that the variations of $\Delta{Q}$ and $\Delta{P}$ with $\Omega$ follow a clear power-law relationship~\cite{gietka2022squeezing}. Linear fitting in logarithmic coordinates demonstrates a strong linear correlation between 
$\ln{\Delta{Q}}$, $\ln{\Delta{P}}$ and $\ln{\Omega}$, confirming the scaling behavior of quantum fluctuations with system size.

To explore the evolution of squeezing characteristics during state preparation with successive blocks, we systematically analyze the variation of $\Delta Q$ and $\Delta P$ with circuit depth for different $\Omega$ (Fig.~\ref{Fig:SPp}(a) and (b)). For a fixed number of blocks
$p=12$, the HVA exhibits stable block-by-block learning behavior. In particular, the number of optimal layers required to reach saturation is proportional to $\ln{\Omega}$, revealing the quantitative impact of effective system size on the complexity of quantum state preparation.

Theoretical analysis indicates that $\Delta{P}<\frac{1}{\sqrt{2}}$ is a key criterion for squeezed states, and our exact numerical results fully satisfy this condition. Additionally, the relation $\Delta{Q}\Delta{P}\approx\frac{1}{2}$ confirms that the system consistently adheres to the fundamental uncertainty principle throughout its evolution. These results provide important quantitative insights into the quantum properties of the QRM critical state and establish a theoretical foundation for subsequent quantum simulation experiments.

\section{conclusion}\label{sec:level4}
%In summary, this study uses a approach for the quantum simulation of continuous-discrete mixed-variable systems and uncovers the squeezing characteristics of the QRM critical states. In the future, this method could be extended to multi-mode quantum systems to explore the effects of decoherence on compressed states
%\newpage

In this work, we have demonstrated the effectiveness of the hybrid-variable VQA in preparing and analyzing the critical states of the QRM. By employing the Hamiltonian variational ansätze with a structured decomposition of the Hamiltonian, we efficiently capture the competition between different quantum fluctuations, enabling accurate preparation of critical states even as the system approaches the thermodynamic limit. The almost linear scaling of the required HVA blocks with the effective system size highlights the algorithm’s efficiency, while the approximate uniform squeezing mechanism across successive blocks provides a clear physical interpretation for this scaling behavior. 

%\textcolor{red}{Similar to the Rabi model, this scaling behavior also holds for certain quantum many-body systems, such as the TFIM~\cite{ho2019efficient,wu2019variational}, the XXZ model~\cite{wiersema2020exploring}, and the XY model \cite{niu2019optimizing}.  When these models are simulated using HVA-based variational quantum algorithms, the required circuit depth exhibits a linear scaling with the system size. In many-body physics, such scaling behavior may be closely related to the structure of quantum entanglement.Specifically, in one-dimensional local Hamiltonians, the ground states typically obey an area law for entanglement entropy, implying that the amount of entanglement between subsystems grows only weakly with system size. As a result, the variational ansätze can efficiently capture the relevant correlations using a circuit whose depth increases linearly with the number of sites. This linear scaling reflects the locality of interactions and the limited range of entanglement propagation during the variational optimization. In contrast, for systems exhibiting volume-law entanglement or long-range correlations, more efficient HVA-based circuits need to be designed}

Our findings not only establish the HVA as a powerful tool for simulating critical states on near-term quantum devices but also offer deeper insights into the structure of these states through the lens of VQA. This approach opens new avenues for exploring quantum critical phenomena in light-matter systems and beyond, suggesting that variational methods with tailored ansätze can serve as both efficient preparators and analytical frameworks for complex quantum states. Future work may extend this methodology to other critical models, further inspiring an understanding of strongly correlated quantum phases with near-term quantum simulators. For instance, the linear scaling behavior of the required circuit depth with the system size also holds for certain quantum many-body systems, such as the TFIM~\cite{ho2019efficient,wu2019variational}, the XXZ model~\cite{wiersema2020exploring}, and the XY model~\cite{niu2019optimizing}. It is possible to understand the behavior by investigating how the Hamiltonian variational ansätze generates spin squeezing layer by layer until it fully satisfies the demanded spin squeezing for the critical state at finite size~\cite{ma2011quantum}. Moreover, other characterizations of quantum critical states, such as entanglement~\cite{Vidal_PRL_2003}, are expected to be useful for understanding complicated quantum critical states under the Hamiltonian variational ansätze as it can constructively tell us how complicated a quantum critical can be.

\begin{acknowledgments}
This work was supported by the National Natural Science Foundation of China (Grant No.12375013) and the Guangdong Basic and Applied Basic Research Fund (Grant No.2023A1515011460).
\end{acknowledgments}

\normalem
\bibliographystyle{apsrev4-2}
\bibliography{n}

\begin{thebibliography}{55}%
\makeatletter
\providecommand \@ifxundefined [1]{%
 \@ifx{#1\undefined}
}%
\providecommand \@ifnum [1]{%
 \ifnum #1\expandafter \@firstoftwo
 \else \expandafter \@secondoftwo
 \fi
}%
\providecommand \@ifx [1]{%
 \ifx #1\expandafter \@firstoftwo
 \else \expandafter \@secondoftwo
 \fi
}%
\providecommand \natexlab [1]{#1}%
\providecommand \enquote  [1]{``#1''}%
\providecommand \bibnamefont  [1]{#1}%
\providecommand \bibfnamefont [1]{#1}%
\providecommand \citenamefont [1]{#1}%
\providecommand \href@noop [0]{\@secondoftwo}%
\providecommand \href [0]{\begingroup \@sanitize@url \@href}%
\providecommand \@href[1]{\@@startlink{#1}\@@href}%
\providecommand \@@href[1]{\endgroup#1\@@endlink}%
\providecommand \@sanitize@url [0]{\catcode `\\12\catcode `\$12\catcode `\&12\catcode `\#12\catcode `\^12\catcode `\_12\catcode `\%12\relax}%
\providecommand \@@startlink[1]{}%
\providecommand \@@endlink[0]{}%
\providecommand \url  [0]{\begingroup\@sanitize@url \@url }%
\providecommand \@url [1]{\endgroup\@href {#1}{\urlprefix }}%
\providecommand \urlprefix  [0]{URL }%
\providecommand \Eprint [0]{\href }%
\providecommand \doibase [0]{https://doi.org/}%
\providecommand \selectlanguage [0]{\@gobble}%
\providecommand \bibinfo  [0]{\@secondoftwo}%
\providecommand \bibfield  [0]{\@secondoftwo}%
\providecommand \translation [1]{[#1]}%
\providecommand \BibitemOpen [0]{}%
\providecommand \bibitemStop [0]{}%
\providecommand \bibitemNoStop [0]{.\EOS\space}%
\providecommand \EOS [0]{\spacefactor3000\relax}%
\providecommand \BibitemShut  [1]{\csname bibitem#1\endcsname}%
\let\auto@bib@innerbib\@empty
%</preamble>
\bibitem [{\citenamefont {Vojta}(2003)}]{vojta2003quantum}%
  \BibitemOpen
  \bibfield  {author} {\bibinfo {author} {\bibfnamefont {M.}~\bibnamefont {Vojta}},\ }\href@noop {} {\bibfield  {journal} {\bibinfo  {journal} {Reports on Progress in Physics}\ }\textbf {\bibinfo {volume} {66}},\ \bibinfo {pages} {2069} (\bibinfo {year} {2003})}\BibitemShut {NoStop}%
\bibitem [{\citenamefont {Heyl}(2018)}]{heyl2018dynamical}%
  \BibitemOpen
  \bibfield  {author} {\bibinfo {author} {\bibfnamefont {M.}~\bibnamefont {Heyl}},\ }\href@noop {} {\bibfield  {journal} {\bibinfo  {journal} {Reports on Progress in Physics}\ }\textbf {\bibinfo {volume} {81}},\ \bibinfo {pages} {054001} (\bibinfo {year} {2018})}\BibitemShut {NoStop}%
\bibitem [{\citenamefont {Bakemeier}\ \emph {et~al.}(2012)\citenamefont {Bakemeier}, \citenamefont {Alvermann},\ and\ \citenamefont {Fehske}}]{bakemeier2012quantum}%
  \BibitemOpen
  \bibfield  {author} {\bibinfo {author} {\bibfnamefont {L.}~\bibnamefont {Bakemeier}}, \bibinfo {author} {\bibfnamefont {A.}~\bibnamefont {Alvermann}},\ and\ \bibinfo {author} {\bibfnamefont {H.}~\bibnamefont {Fehske}},\ }\href@noop {} {\bibfield  {journal} {\bibinfo  {journal} {Physical Review A—Atomic, Molecular, and Optical Physics}\ }\textbf {\bibinfo {volume} {85}},\ \bibinfo {pages} {043821} (\bibinfo {year} {2012})}\BibitemShut {NoStop}%
\bibitem [{\citenamefont {Puebla}\ \emph {et~al.}(2013)\citenamefont {Puebla}, \citenamefont {Relano},\ and\ \citenamefont {Retamosa}}]{puebla2013excited}%
  \BibitemOpen
  \bibfield  {author} {\bibinfo {author} {\bibfnamefont {R.}~\bibnamefont {Puebla}}, \bibinfo {author} {\bibfnamefont {A.}~\bibnamefont {Relano}},\ and\ \bibinfo {author} {\bibfnamefont {J.}~\bibnamefont {Retamosa}},\ }\href@noop {} {\bibfield  {journal} {\bibinfo  {journal} {Physical Review A—Atomic, Molecular, and Optical Physics}\ }\textbf {\bibinfo {volume} {87}},\ \bibinfo {pages} {023819} (\bibinfo {year} {2013})}\BibitemShut {NoStop}%
\bibitem [{\citenamefont {Hwang}\ and\ \citenamefont {Plenio}(2016)}]{hwang2016quantum}%
  \BibitemOpen
  \bibfield  {author} {\bibinfo {author} {\bibfnamefont {M.-J.}\ \bibnamefont {Hwang}}\ and\ \bibinfo {author} {\bibfnamefont {M.~B.}\ \bibnamefont {Plenio}},\ }\href@noop {} {\bibfield  {journal} {\bibinfo  {journal} {Physical Review Letters}\ }\textbf {\bibinfo {volume} {117}},\ \bibinfo {pages} {123602} (\bibinfo {year} {2016})}\BibitemShut {NoStop}%
\bibitem [{\citenamefont {Ashhab}\ and\ \citenamefont {Nori}(2010)}]{ashhab2010qubit}%
  \BibitemOpen
  \bibfield  {author} {\bibinfo {author} {\bibfnamefont {S.}~\bibnamefont {Ashhab}}\ and\ \bibinfo {author} {\bibfnamefont {F.}~\bibnamefont {Nori}},\ }\href@noop {} {\bibfield  {journal} {\bibinfo  {journal} {Physical Review A—Atomic, Molecular, and Optical Physics}\ }\textbf {\bibinfo {volume} {81}},\ \bibinfo {pages} {042311} (\bibinfo {year} {2010})}\BibitemShut {NoStop}%
\bibitem [{\citenamefont {Ashhab}(2013)}]{ashhab2013superradiance}%
  \BibitemOpen
  \bibfield  {author} {\bibinfo {author} {\bibfnamefont {S.}~\bibnamefont {Ashhab}},\ }\href@noop {} {\bibfield  {journal} {\bibinfo  {journal} {Physical Review A—Atomic, Molecular, and Optical Physics}\ }\textbf {\bibinfo {volume} {87}},\ \bibinfo {pages} {013826} (\bibinfo {year} {2013})}\BibitemShut {NoStop}%
\bibitem [{\citenamefont {Puebla}\ \emph {et~al.}(2017)\citenamefont {Puebla}, \citenamefont {Hwang}, \citenamefont {Casanova},\ and\ \citenamefont {Plenio}}]{puebla2017probing}%
  \BibitemOpen
  \bibfield  {author} {\bibinfo {author} {\bibfnamefont {R.}~\bibnamefont {Puebla}}, \bibinfo {author} {\bibfnamefont {M.-J.}\ \bibnamefont {Hwang}}, \bibinfo {author} {\bibfnamefont {J.}~\bibnamefont {Casanova}},\ and\ \bibinfo {author} {\bibfnamefont {M.~B.}\ \bibnamefont {Plenio}},\ }\href@noop {} {\bibfield  {journal} {\bibinfo  {journal} {Physical review letters}\ }\textbf {\bibinfo {volume} {118}},\ \bibinfo {pages} {073001} (\bibinfo {year} {2017})}\BibitemShut {NoStop}%
\bibitem [{\citenamefont {Hwang}\ \emph {et~al.}(2015)\citenamefont {Hwang}, \citenamefont {Puebla},\ and\ \citenamefont {Plenio}}]{hwang2015quantum}%
  \BibitemOpen
  \bibfield  {author} {\bibinfo {author} {\bibfnamefont {M.-J.}\ \bibnamefont {Hwang}}, \bibinfo {author} {\bibfnamefont {R.}~\bibnamefont {Puebla}},\ and\ \bibinfo {author} {\bibfnamefont {M.~B.}\ \bibnamefont {Plenio}},\ }\href@noop {} {\bibfield  {journal} {\bibinfo  {journal} {Physical review letters}\ }\textbf {\bibinfo {volume} {115}},\ \bibinfo {pages} {180404} (\bibinfo {year} {2015})}\BibitemShut {NoStop}%
\bibitem [{\citenamefont {Puebla}\ \emph {et~al.}(2016)\citenamefont {Puebla}, \citenamefont {Hwang},\ and\ \citenamefont {Plenio}}]{puebla2016excited}%
  \BibitemOpen
  \bibfield  {author} {\bibinfo {author} {\bibfnamefont {R.}~\bibnamefont {Puebla}}, \bibinfo {author} {\bibfnamefont {M.-J.}\ \bibnamefont {Hwang}},\ and\ \bibinfo {author} {\bibfnamefont {M.~B.}\ \bibnamefont {Plenio}},\ }\href@noop {} {\bibfield  {journal} {\bibinfo  {journal} {Physical Review A}\ }\textbf {\bibinfo {volume} {94}},\ \bibinfo {pages} {023835} (\bibinfo {year} {2016})}\BibitemShut {NoStop}%
\bibitem [{\citenamefont {Chen}\ \emph {et~al.}(2024{\natexlab{a}})\citenamefont {Chen}, \citenamefont {Qiu}, \citenamefont {Miranowicz}, \citenamefont {Lambert}, \citenamefont {Qin}, \citenamefont {Stassi}, \citenamefont {Xia}, \citenamefont {Zheng},\ and\ \citenamefont {Nori}}]{chen2024sudden}%
  \BibitemOpen
  \bibfield  {author} {\bibinfo {author} {\bibfnamefont {Y.-H.}\ \bibnamefont {Chen}}, \bibinfo {author} {\bibfnamefont {Y.}~\bibnamefont {Qiu}}, \bibinfo {author} {\bibfnamefont {A.}~\bibnamefont {Miranowicz}}, \bibinfo {author} {\bibfnamefont {N.}~\bibnamefont {Lambert}}, \bibinfo {author} {\bibfnamefont {W.}~\bibnamefont {Qin}}, \bibinfo {author} {\bibfnamefont {R.}~\bibnamefont {Stassi}}, \bibinfo {author} {\bibfnamefont {Y.}~\bibnamefont {Xia}}, \bibinfo {author} {\bibfnamefont {S.-B.}\ \bibnamefont {Zheng}},\ and\ \bibinfo {author} {\bibfnamefont {F.}~\bibnamefont {Nori}},\ }\href@noop {} {\bibfield  {journal} {\bibinfo  {journal} {Communications Physics}\ }\textbf {\bibinfo {volume} {7}},\ \bibinfo {pages} {5} (\bibinfo {year} {2024}{\natexlab{a}})}\BibitemShut {NoStop}%
\bibitem [{\citenamefont {Chen}\ \emph {et~al.}(2024{\natexlab{b}})\citenamefont {Chen}, \citenamefont {Shi}, \citenamefont {Nori},\ and\ \citenamefont {Xia}}]{chen2024error}%
  \BibitemOpen
  \bibfield  {author} {\bibinfo {author} {\bibfnamefont {Y.-H.}\ \bibnamefont {Chen}}, \bibinfo {author} {\bibfnamefont {Z.-C.}\ \bibnamefont {Shi}}, \bibinfo {author} {\bibfnamefont {F.}~\bibnamefont {Nori}},\ and\ \bibinfo {author} {\bibfnamefont {Y.}~\bibnamefont {Xia}},\ }\href@noop {} {\bibfield  {journal} {\bibinfo  {journal} {Physical Review Letters}\ }\textbf {\bibinfo {volume} {133}},\ \bibinfo {pages} {033603} (\bibinfo {year} {2024}{\natexlab{b}})}\BibitemShut {NoStop}%
\bibitem [{\citenamefont {Liu}\ \emph {et~al.}(2022)\citenamefont {Liu}, \citenamefont {Zhao}, \citenamefont {Yang},\ and\ \citenamefont {Luo}}]{liu2022process}%
  \BibitemOpen
  \bibfield  {author} {\bibinfo {author} {\bibfnamefont {J.}~\bibnamefont {Liu}}, \bibinfo {author} {\bibfnamefont {M.}~\bibnamefont {Zhao}}, \bibinfo {author} {\bibfnamefont {Y.-T.}\ \bibnamefont {Yang}},\ and\ \bibinfo {author} {\bibfnamefont {H.-G.}\ \bibnamefont {Luo}},\ }\href@noop {} {\bibfield  {journal} {\bibinfo  {journal} {arXiv preprint arXiv:2211.16233}\ } (\bibinfo {year} {2022})}\BibitemShut {NoStop}%
\bibitem [{\citenamefont {Gietka}\ \emph {et~al.}(2023)\citenamefont {Gietka}, \citenamefont {Hotter},\ and\ \citenamefont {Ritsch}}]{gietka2023unique}%
  \BibitemOpen
  \bibfield  {author} {\bibinfo {author} {\bibfnamefont {K.}~\bibnamefont {Gietka}}, \bibinfo {author} {\bibfnamefont {C.}~\bibnamefont {Hotter}},\ and\ \bibinfo {author} {\bibfnamefont {H.}~\bibnamefont {Ritsch}},\ }\href@noop {} {\bibfield  {journal} {\bibinfo  {journal} {Physical Review Letters}\ }\textbf {\bibinfo {volume} {131}},\ \bibinfo {pages} {223604} (\bibinfo {year} {2023})}\BibitemShut {NoStop}%
\bibitem [{\citenamefont {Feynman}(2018)}]{feynman2018simulating}%
  \BibitemOpen
  \bibfield  {author} {\bibinfo {author} {\bibfnamefont {R.~P.}\ \bibnamefont {Feynman}},\ }in\ \href@noop {} {\emph {\bibinfo {booktitle} {Feynman and computation}}}\ (\bibinfo  {publisher} {cRc Press},\ \bibinfo {year} {2018})\ pp.\ \bibinfo {pages} {133--153}\BibitemShut {NoStop}%
\bibitem [{\citenamefont {Lloyd}(1996)}]{lloyd1996universal}%
  \BibitemOpen
  \bibfield  {author} {\bibinfo {author} {\bibfnamefont {S.}~\bibnamefont {Lloyd}},\ }\href@noop {} {\bibfield  {journal} {\bibinfo  {journal} {Science}\ }\textbf {\bibinfo {volume} {273}},\ \bibinfo {pages} {1073} (\bibinfo {year} {1996})}\BibitemShut {NoStop}%
\bibitem [{\citenamefont {Greiner}\ \emph {et~al.}(2002)\citenamefont {Greiner}, \citenamefont {Mandel}, \citenamefont {Esslinger}, \citenamefont {H{\"a}nsch},\ and\ \citenamefont {Bloch}}]{greiner2002quantum}%
  \BibitemOpen
  \bibfield  {author} {\bibinfo {author} {\bibfnamefont {M.}~\bibnamefont {Greiner}}, \bibinfo {author} {\bibfnamefont {O.}~\bibnamefont {Mandel}}, \bibinfo {author} {\bibfnamefont {T.}~\bibnamefont {Esslinger}}, \bibinfo {author} {\bibfnamefont {T.~W.}\ \bibnamefont {H{\"a}nsch}},\ and\ \bibinfo {author} {\bibfnamefont {I.}~\bibnamefont {Bloch}},\ }\href@noop {} {\bibfield  {journal} {\bibinfo  {journal} {nature}\ }\textbf {\bibinfo {volume} {415}},\ \bibinfo {pages} {39} (\bibinfo {year} {2002})}\BibitemShut {NoStop}%
\bibitem [{\citenamefont {Blatt}\ and\ \citenamefont {Roos}(2012)}]{blatt2012quantum}%
  \BibitemOpen
  \bibfield  {author} {\bibinfo {author} {\bibfnamefont {R.}~\bibnamefont {Blatt}}\ and\ \bibinfo {author} {\bibfnamefont {C.~F.}\ \bibnamefont {Roos}},\ }\href@noop {} {\bibfield  {journal} {\bibinfo  {journal} {Nature Physics}\ }\textbf {\bibinfo {volume} {8}},\ \bibinfo {pages} {277} (\bibinfo {year} {2012})}\BibitemShut {NoStop}%
\bibitem [{\citenamefont {Ebadi}\ \emph {et~al.}(2021)\citenamefont {Ebadi}, \citenamefont {Wang}, \citenamefont {Levine}, \citenamefont {Keesling}, \citenamefont {Semeghini}, \citenamefont {Omran}, \citenamefont {Bluvstein}, \citenamefont {Samajdar}, \citenamefont {Pichler}, \citenamefont {Ho} \emph {et~al.}}]{ebadi2021quantum}%
  \BibitemOpen
  \bibfield  {author} {\bibinfo {author} {\bibfnamefont {S.}~\bibnamefont {Ebadi}}, \bibinfo {author} {\bibfnamefont {T.~T.}\ \bibnamefont {Wang}}, \bibinfo {author} {\bibfnamefont {H.}~\bibnamefont {Levine}}, \bibinfo {author} {\bibfnamefont {A.}~\bibnamefont {Keesling}}, \bibinfo {author} {\bibfnamefont {G.}~\bibnamefont {Semeghini}}, \bibinfo {author} {\bibfnamefont {A.}~\bibnamefont {Omran}}, \bibinfo {author} {\bibfnamefont {D.}~\bibnamefont {Bluvstein}}, \bibinfo {author} {\bibfnamefont {R.}~\bibnamefont {Samajdar}}, \bibinfo {author} {\bibfnamefont {H.}~\bibnamefont {Pichler}}, \bibinfo {author} {\bibfnamefont {W.~W.}\ \bibnamefont {Ho}}, \emph {et~al.},\ }\href@noop {} {\bibfield  {journal} {\bibinfo  {journal} {Nature}\ }\textbf {\bibinfo {volume} {595}},\ \bibinfo {pages} {227} (\bibinfo {year} {2021})}\BibitemShut {NoStop}%
\bibitem [{\citenamefont {Cerezo}\ \emph {et~al.}(2021)\citenamefont {Cerezo}, \citenamefont {Arrasmith}, \citenamefont {Babbush}, \citenamefont {Benjamin}, \citenamefont {Endo}, \citenamefont {Fujii}, \citenamefont {McClean}, \citenamefont {Mitarai}, \citenamefont {Yuan}, \citenamefont {Cincio} \emph {et~al.}}]{cerezo2021variational}%
  \BibitemOpen
  \bibfield  {author} {\bibinfo {author} {\bibfnamefont {M.}~\bibnamefont {Cerezo}}, \bibinfo {author} {\bibfnamefont {A.}~\bibnamefont {Arrasmith}}, \bibinfo {author} {\bibfnamefont {R.}~\bibnamefont {Babbush}}, \bibinfo {author} {\bibfnamefont {S.~C.}\ \bibnamefont {Benjamin}}, \bibinfo {author} {\bibfnamefont {S.}~\bibnamefont {Endo}}, \bibinfo {author} {\bibfnamefont {K.}~\bibnamefont {Fujii}}, \bibinfo {author} {\bibfnamefont {J.~R.}\ \bibnamefont {McClean}}, \bibinfo {author} {\bibfnamefont {K.}~\bibnamefont {Mitarai}}, \bibinfo {author} {\bibfnamefont {X.}~\bibnamefont {Yuan}}, \bibinfo {author} {\bibfnamefont {L.}~\bibnamefont {Cincio}}, \emph {et~al.},\ }\href@noop {} {\bibfield  {journal} {\bibinfo  {journal} {Nature Reviews Physics}\ }\textbf {\bibinfo {volume} {3}},\ \bibinfo {pages} {625} (\bibinfo {year} {2021})}\BibitemShut {NoStop}%
\bibitem [{\citenamefont {Preskill}(2018)}]{preskill2018quantum}%
  \BibitemOpen
  \bibfield  {author} {\bibinfo {author} {\bibfnamefont {J.}~\bibnamefont {Preskill}},\ }\href@noop {} {\bibfield  {journal} {\bibinfo  {journal} {Quantum}\ }\textbf {\bibinfo {volume} {2}},\ \bibinfo {pages} {79} (\bibinfo {year} {2018})}\BibitemShut {NoStop}%
\bibitem [{\citenamefont {Bharti}\ \emph {et~al.}(2022)\citenamefont {Bharti}, \citenamefont {Cervera-Lierta}, \citenamefont {Kyaw}, \citenamefont {Haug}, \citenamefont {Alperin-Lea}, \citenamefont {Anand}, \citenamefont {Degroote}, \citenamefont {Heimonen}, \citenamefont {Kottmann}, \citenamefont {Menke} \emph {et~al.}}]{bharti2022noisy}%
  \BibitemOpen
  \bibfield  {author} {\bibinfo {author} {\bibfnamefont {K.}~\bibnamefont {Bharti}}, \bibinfo {author} {\bibfnamefont {A.}~\bibnamefont {Cervera-Lierta}}, \bibinfo {author} {\bibfnamefont {T.~H.}\ \bibnamefont {Kyaw}}, \bibinfo {author} {\bibfnamefont {T.}~\bibnamefont {Haug}}, \bibinfo {author} {\bibfnamefont {S.}~\bibnamefont {Alperin-Lea}}, \bibinfo {author} {\bibfnamefont {A.}~\bibnamefont {Anand}}, \bibinfo {author} {\bibfnamefont {M.}~\bibnamefont {Degroote}}, \bibinfo {author} {\bibfnamefont {H.}~\bibnamefont {Heimonen}}, \bibinfo {author} {\bibfnamefont {J.~S.}\ \bibnamefont {Kottmann}}, \bibinfo {author} {\bibfnamefont {T.}~\bibnamefont {Menke}}, \emph {et~al.},\ }\href@noop {} {\bibfield  {journal} {\bibinfo  {journal} {Reviews of Modern Physics}\ }\textbf {\bibinfo {volume} {94}},\ \bibinfo {pages} {015004} (\bibinfo {year} {2022})}\BibitemShut {NoStop}%
\bibitem [{\citenamefont {Farhi}\ \emph {et~al.}(2014)\citenamefont {Farhi}, \citenamefont {Goldstone},\ and\ \citenamefont {Gutmann}}]{farhi2014quantum}%
  \BibitemOpen
  \bibfield  {author} {\bibinfo {author} {\bibfnamefont {E.}~\bibnamefont {Farhi}}, \bibinfo {author} {\bibfnamefont {J.}~\bibnamefont {Goldstone}},\ and\ \bibinfo {author} {\bibfnamefont {S.}~\bibnamefont {Gutmann}},\ }\href@noop {} {\bibfield  {journal} {\bibinfo  {journal} {arXiv preprint arXiv:1411.4028}\ } (\bibinfo {year} {2014})}\BibitemShut {NoStop}%
\bibitem [{\citenamefont {Farhi}\ and\ \citenamefont {Harrow}(2016)}]{farhi2016quantum}%
  \BibitemOpen
  \bibfield  {author} {\bibinfo {author} {\bibfnamefont {E.}~\bibnamefont {Farhi}}\ and\ \bibinfo {author} {\bibfnamefont {A.~W.}\ \bibnamefont {Harrow}},\ }\href@noop {} {\bibfield  {journal} {\bibinfo  {journal} {arXiv preprint arXiv:1602.07674}\ } (\bibinfo {year} {2016})}\BibitemShut {NoStop}%
\bibitem [{\citenamefont {Peruzzo}\ \emph {et~al.}(2014)\citenamefont {Peruzzo}, \citenamefont {McClean}, \citenamefont {Shadbolt}, \citenamefont {Yung}, \citenamefont {Zhou}, \citenamefont {Love}, \citenamefont {Aspuru-Guzik},\ and\ \citenamefont {O’brien}}]{peruzzo2014variational}%
  \BibitemOpen
  \bibfield  {author} {\bibinfo {author} {\bibfnamefont {A.}~\bibnamefont {Peruzzo}}, \bibinfo {author} {\bibfnamefont {J.}~\bibnamefont {McClean}}, \bibinfo {author} {\bibfnamefont {P.}~\bibnamefont {Shadbolt}}, \bibinfo {author} {\bibfnamefont {M.-H.}\ \bibnamefont {Yung}}, \bibinfo {author} {\bibfnamefont {X.-Q.}\ \bibnamefont {Zhou}}, \bibinfo {author} {\bibfnamefont {P.~J.}\ \bibnamefont {Love}}, \bibinfo {author} {\bibfnamefont {A.}~\bibnamefont {Aspuru-Guzik}},\ and\ \bibinfo {author} {\bibfnamefont {J.~L.}\ \bibnamefont {O’brien}},\ }\href@noop {} {\bibfield  {journal} {\bibinfo  {journal} {Nature communications}\ }\textbf {\bibinfo {volume} {5}},\ \bibinfo {pages} {4213} (\bibinfo {year} {2014})}\BibitemShut {NoStop}%
\bibitem [{\citenamefont {Nakanishi}\ \emph {et~al.}(2019)\citenamefont {Nakanishi}, \citenamefont {Mitarai},\ and\ \citenamefont {Fujii}}]{nakanishi2019subspace}%
  \BibitemOpen
  \bibfield  {author} {\bibinfo {author} {\bibfnamefont {K.~M.}\ \bibnamefont {Nakanishi}}, \bibinfo {author} {\bibfnamefont {K.}~\bibnamefont {Mitarai}},\ and\ \bibinfo {author} {\bibfnamefont {K.}~\bibnamefont {Fujii}},\ }\href@noop {} {\bibfield  {journal} {\bibinfo  {journal} {Physical Review Research}\ }\textbf {\bibinfo {volume} {1}},\ \bibinfo {pages} {033062} (\bibinfo {year} {2019})}\BibitemShut {NoStop}%
\bibitem [{\citenamefont {Wecker}\ \emph {et~al.}(2015)\citenamefont {Wecker}, \citenamefont {Hastings},\ and\ \citenamefont {Troyer}}]{wecker2015progress}%
  \BibitemOpen
  \bibfield  {author} {\bibinfo {author} {\bibfnamefont {D.}~\bibnamefont {Wecker}}, \bibinfo {author} {\bibfnamefont {M.~B.}\ \bibnamefont {Hastings}},\ and\ \bibinfo {author} {\bibfnamefont {M.}~\bibnamefont {Troyer}},\ }\href@noop {} {\bibfield  {journal} {\bibinfo  {journal} {Physical Review A}\ }\textbf {\bibinfo {volume} {92}},\ \bibinfo {pages} {042303} (\bibinfo {year} {2015})}\BibitemShut {NoStop}%
\bibitem [{\citenamefont {Kokail}\ \emph {et~al.}(2019)\citenamefont {Kokail}, \citenamefont {Maier}, \citenamefont {van Bijnen}, \citenamefont {Brydges}, \citenamefont {Joshi}, \citenamefont {Jurcevic}, \citenamefont {Muschik}, \citenamefont {Silvi}, \citenamefont {Blatt}, \citenamefont {Roos} \emph {et~al.}}]{kokail2019self}%
  \BibitemOpen
  \bibfield  {author} {\bibinfo {author} {\bibfnamefont {C.}~\bibnamefont {Kokail}}, \bibinfo {author} {\bibfnamefont {C.}~\bibnamefont {Maier}}, \bibinfo {author} {\bibfnamefont {R.}~\bibnamefont {van Bijnen}}, \bibinfo {author} {\bibfnamefont {T.}~\bibnamefont {Brydges}}, \bibinfo {author} {\bibfnamefont {M.~K.}\ \bibnamefont {Joshi}}, \bibinfo {author} {\bibfnamefont {P.}~\bibnamefont {Jurcevic}}, \bibinfo {author} {\bibfnamefont {C.~A.}\ \bibnamefont {Muschik}}, \bibinfo {author} {\bibfnamefont {P.}~\bibnamefont {Silvi}}, \bibinfo {author} {\bibfnamefont {R.}~\bibnamefont {Blatt}}, \bibinfo {author} {\bibfnamefont {C.~F.}\ \bibnamefont {Roos}}, \emph {et~al.},\ }\href@noop {} {\bibfield  {journal} {\bibinfo  {journal} {Nature}\ }\textbf {\bibinfo {volume} {569}},\ \bibinfo {pages} {355} (\bibinfo {year} {2019})}\BibitemShut {NoStop}%
\bibitem [{\citenamefont {Evenbly}\ and\ \citenamefont {Vidal}(2013)}]{evenbly2013quantum}%
  \BibitemOpen
  \bibfield  {author} {\bibinfo {author} {\bibfnamefont {G.}~\bibnamefont {Evenbly}}\ and\ \bibinfo {author} {\bibfnamefont {G.}~\bibnamefont {Vidal}},\ }in\ \href@noop {} {\emph {\bibinfo {booktitle} {Strongly correlated systems: numerical methods}}}\ (\bibinfo  {publisher} {Springer},\ \bibinfo {year} {2013})\ pp.\ \bibinfo {pages} {99--130}\BibitemShut {NoStop}%
\bibitem [{\citenamefont {Arg{\"u}ello-Luengo}\ \emph {et~al.}(2022)\citenamefont {Arg{\"u}ello-Luengo}, \citenamefont {Milsted},\ and\ \citenamefont {Vidal}}]{arguello2022generalized}%
  \BibitemOpen
  \bibfield  {author} {\bibinfo {author} {\bibfnamefont {J.}~\bibnamefont {Arg{\"u}ello-Luengo}}, \bibinfo {author} {\bibfnamefont {A.}~\bibnamefont {Milsted}},\ and\ \bibinfo {author} {\bibfnamefont {G.}~\bibnamefont {Vidal}},\ }\href@noop {} {\bibfield  {journal} {\bibinfo  {journal} {arXiv preprint arXiv:2212.06740}\ } (\bibinfo {year} {2022})}\BibitemShut {NoStop}%
\bibitem [{\citenamefont {Luchnikov}\ \emph {et~al.}(2021)\citenamefont {Luchnikov}, \citenamefont {Berezutskii},\ and\ \citenamefont {Fedorov}}]{luchnikov2021simulating}%
  \BibitemOpen
  \bibfield  {author} {\bibinfo {author} {\bibfnamefont {I.}~\bibnamefont {Luchnikov}}, \bibinfo {author} {\bibfnamefont {A.}~\bibnamefont {Berezutskii}},\ and\ \bibinfo {author} {\bibfnamefont {A.}~\bibnamefont {Fedorov}},\ }\href@noop {} {\bibfield  {journal} {\bibinfo  {journal} {arXiv preprint arXiv:2112.14046}\ } (\bibinfo {year} {2021})}\BibitemShut {NoStop}%
\bibitem [{\citenamefont {Ho}\ and\ \citenamefont {Hsieh}(2019)}]{ho2019efficient}%
  \BibitemOpen
  \bibfield  {author} {\bibinfo {author} {\bibfnamefont {W.~W.}\ \bibnamefont {Ho}}\ and\ \bibinfo {author} {\bibfnamefont {T.~H.}\ \bibnamefont {Hsieh}},\ }\href@noop {} {\bibfield  {journal} {\bibinfo  {journal} {SciPost Physics}\ }\textbf {\bibinfo {volume} {6}},\ \bibinfo {pages} {029} (\bibinfo {year} {2019})}\BibitemShut {NoStop}%
\bibitem [{\citenamefont {Lloyd}(2003)}]{lloyd2003hybrid}%
  \BibitemOpen
  \bibfield  {author} {\bibinfo {author} {\bibfnamefont {S.}~\bibnamefont {Lloyd}},\ }in\ \href@noop {} {\emph {\bibinfo {booktitle} {Quantum information with continuous variables}}}\ (\bibinfo  {publisher} {Springer},\ \bibinfo {year} {2003})\ pp.\ \bibinfo {pages} {37--45}\BibitemShut {NoStop}%
\bibitem [{\citenamefont {Lloyd}\ and\ \citenamefont {Braunstein}(1999)}]{lloyd1999quantum}%
  \BibitemOpen
  \bibfield  {author} {\bibinfo {author} {\bibfnamefont {S.}~\bibnamefont {Lloyd}}\ and\ \bibinfo {author} {\bibfnamefont {S.~L.}\ \bibnamefont {Braunstein}},\ }\href@noop {} {\bibfield  {journal} {\bibinfo  {journal} {Physical Review Letters}\ }\textbf {\bibinfo {volume} {82}},\ \bibinfo {pages} {1784} (\bibinfo {year} {1999})}\BibitemShut {NoStop}%
\bibitem [{\citenamefont {Zhang}\ \emph {et~al.}(2021)\citenamefont {Zhang}, \citenamefont {Zhang}, \citenamefont {Xue}, \citenamefont {Zhu},\ and\ \citenamefont {Wang}}]{zhang2021continuous}%
  \BibitemOpen
  \bibfield  {author} {\bibinfo {author} {\bibfnamefont {D.-B.}\ \bibnamefont {Zhang}}, \bibinfo {author} {\bibfnamefont {G.-Q.}\ \bibnamefont {Zhang}}, \bibinfo {author} {\bibfnamefont {Z.-Y.}\ \bibnamefont {Xue}}, \bibinfo {author} {\bibfnamefont {S.-L.}\ \bibnamefont {Zhu}},\ and\ \bibinfo {author} {\bibfnamefont {Z.}~\bibnamefont {Wang}},\ }\href@noop {} {\bibfield  {journal} {\bibinfo  {journal} {Physical review letters}\ }\textbf {\bibinfo {volume} {127}},\ \bibinfo {pages} {020502} (\bibinfo {year} {2021})}\BibitemShut {NoStop}%
\bibitem [{\citenamefont {Wang}\ \emph {et~al.}(2025)\citenamefont {Wang}, \citenamefont {Xie}, \citenamefont {Liu}, \citenamefont {Song}, \citenamefont {Xiong},\ and\ \citenamefont {Wang}}]{wang2025quantum}%
  \BibitemOpen
  \bibfield  {author} {\bibinfo {author} {\bibfnamefont {D.}~\bibnamefont {Wang}}, \bibinfo {author} {\bibfnamefont {L.}~\bibnamefont {Xie}}, \bibinfo {author} {\bibfnamefont {J.}~\bibnamefont {Liu}}, \bibinfo {author} {\bibfnamefont {Y.}~\bibnamefont {Song}}, \bibinfo {author} {\bibfnamefont {W.}~\bibnamefont {Xiong}},\ and\ \bibinfo {author} {\bibfnamefont {M.}~\bibnamefont {Wang}},\ }\href@noop {} {\bibfield  {journal} {\bibinfo  {journal} {Physical Review A}\ }\textbf {\bibinfo {volume} {111}},\ \bibinfo {pages} {012438} (\bibinfo {year} {2025})}\BibitemShut {NoStop}%
\bibitem [{\citenamefont {Dutta}\ \emph {et~al.}(2025)\citenamefont {Dutta}, \citenamefont {Allen}, \citenamefont {Vu}, \citenamefont {Xu}, \citenamefont {Liu}, \citenamefont {Miao}, \citenamefont {Wang}, \citenamefont {Surana}, \citenamefont {Wang}, \citenamefont {Ding} \emph {et~al.}}]{dutta2025solving}%
  \BibitemOpen
  \bibfield  {author} {\bibinfo {author} {\bibfnamefont {R.}~\bibnamefont {Dutta}}, \bibinfo {author} {\bibfnamefont {B.}~\bibnamefont {Allen}}, \bibinfo {author} {\bibfnamefont {N.~P.}\ \bibnamefont {Vu}}, \bibinfo {author} {\bibfnamefont {C.}~\bibnamefont {Xu}}, \bibinfo {author} {\bibfnamefont {K.}~\bibnamefont {Liu}}, \bibinfo {author} {\bibfnamefont {F.}~\bibnamefont {Miao}}, \bibinfo {author} {\bibfnamefont {B.}~\bibnamefont {Wang}}, \bibinfo {author} {\bibfnamefont {A.}~\bibnamefont {Surana}}, \bibinfo {author} {\bibfnamefont {C.}~\bibnamefont {Wang}}, \bibinfo {author} {\bibfnamefont {Y.}~\bibnamefont {Ding}}, \emph {et~al.},\ }\href@noop {} {\bibfield  {journal} {\bibinfo  {journal} {arXiv preprint arXiv:2501.11735}\ } (\bibinfo {year} {2025})}\BibitemShut {NoStop}%
\bibitem [{\citenamefont {Araz}\ \emph {et~al.}(2024)\citenamefont {Araz}, \citenamefont {Grau}, \citenamefont {Montgomery},\ and\ \citenamefont {Ringer}}]{araz2024toward}%
  \BibitemOpen
  \bibfield  {author} {\bibinfo {author} {\bibfnamefont {J.~Y.}\ \bibnamefont {Araz}}, \bibinfo {author} {\bibfnamefont {M.}~\bibnamefont {Grau}}, \bibinfo {author} {\bibfnamefont {J.}~\bibnamefont {Montgomery}},\ and\ \bibinfo {author} {\bibfnamefont {F.}~\bibnamefont {Ringer}},\ }\href@noop {} {\bibfield  {journal} {\bibinfo  {journal} {arXiv preprint arXiv:2410.07346}\ } (\bibinfo {year} {2024})}\BibitemShut {NoStop}%
\bibitem [{\citenamefont {Braak}(2011)}]{braak2011integrability}%
  \BibitemOpen
  \bibfield  {author} {\bibinfo {author} {\bibfnamefont {D.}~\bibnamefont {Braak}},\ }\href@noop {} {\bibfield  {journal} {\bibinfo  {journal} {Physical Review Letters}\ }\textbf {\bibinfo {volume} {107}},\ \bibinfo {pages} {100401} (\bibinfo {year} {2011})}\BibitemShut {NoStop}%
\bibitem [{\citenamefont {Andersen}\ \emph {et~al.}(2015)\citenamefont {Andersen}, \citenamefont {Neergaard-Nielsen}, \citenamefont {Van~Loock},\ and\ \citenamefont {Furusawa}}]{andersen2015hybrid}%
  \BibitemOpen
  \bibfield  {author} {\bibinfo {author} {\bibfnamefont {U.~L.}\ \bibnamefont {Andersen}}, \bibinfo {author} {\bibfnamefont {J.~S.}\ \bibnamefont {Neergaard-Nielsen}}, \bibinfo {author} {\bibfnamefont {P.}~\bibnamefont {Van~Loock}},\ and\ \bibinfo {author} {\bibfnamefont {A.}~\bibnamefont {Furusawa}},\ }\href@noop {} {\bibfield  {journal} {\bibinfo  {journal} {Nature Physics}\ }\textbf {\bibinfo {volume} {11}},\ \bibinfo {pages} {713} (\bibinfo {year} {2015})}\BibitemShut {NoStop}%
\bibitem [{\citenamefont {Wallraff}\ \emph {et~al.}(2004)\citenamefont {Wallraff}, \citenamefont {Schuster}, \citenamefont {Blais}, \citenamefont {Frunzio}, \citenamefont {Huang}, \citenamefont {Majer}, \citenamefont {Kumar}, \citenamefont {Girvin},\ and\ \citenamefont {Schoelkopf}}]{wallraff2004strong}%
  \BibitemOpen
  \bibfield  {author} {\bibinfo {author} {\bibfnamefont {A.}~\bibnamefont {Wallraff}}, \bibinfo {author} {\bibfnamefont {D.~I.}\ \bibnamefont {Schuster}}, \bibinfo {author} {\bibfnamefont {A.}~\bibnamefont {Blais}}, \bibinfo {author} {\bibfnamefont {L.}~\bibnamefont {Frunzio}}, \bibinfo {author} {\bibfnamefont {R.-S.}\ \bibnamefont {Huang}}, \bibinfo {author} {\bibfnamefont {J.}~\bibnamefont {Majer}}, \bibinfo {author} {\bibfnamefont {S.}~\bibnamefont {Kumar}}, \bibinfo {author} {\bibfnamefont {S.~M.}\ \bibnamefont {Girvin}},\ and\ \bibinfo {author} {\bibfnamefont {R.~J.}\ \bibnamefont {Schoelkopf}},\ }\href@noop {} {\bibfield  {journal} {\bibinfo  {journal} {Nature}\ }\textbf {\bibinfo {volume} {431}},\ \bibinfo {pages} {162} (\bibinfo {year} {2004})}\BibitemShut {NoStop}%
\bibitem [{\citenamefont {Paik}\ \emph {et~al.}(2011)\citenamefont {Paik}, \citenamefont {Schuster}, \citenamefont {Bishop}, \citenamefont {Kirchmair}, \citenamefont {Catelani}, \citenamefont {Sears}, \citenamefont {Johnson}, \citenamefont {Reagor}, \citenamefont {Frunzio}, \citenamefont {Glazman} \emph {et~al.}}]{paik2011observation}%
  \BibitemOpen
  \bibfield  {author} {\bibinfo {author} {\bibfnamefont {H.}~\bibnamefont {Paik}}, \bibinfo {author} {\bibfnamefont {D.~I.}\ \bibnamefont {Schuster}}, \bibinfo {author} {\bibfnamefont {L.~S.}\ \bibnamefont {Bishop}}, \bibinfo {author} {\bibfnamefont {G.}~\bibnamefont {Kirchmair}}, \bibinfo {author} {\bibfnamefont {G.}~\bibnamefont {Catelani}}, \bibinfo {author} {\bibfnamefont {A.~P.}\ \bibnamefont {Sears}}, \bibinfo {author} {\bibfnamefont {B.}~\bibnamefont {Johnson}}, \bibinfo {author} {\bibfnamefont {M.}~\bibnamefont {Reagor}}, \bibinfo {author} {\bibfnamefont {L.}~\bibnamefont {Frunzio}}, \bibinfo {author} {\bibfnamefont {L.~I.}\ \bibnamefont {Glazman}}, \emph {et~al.},\ }\href@noop {} {\bibfield  {journal} {\bibinfo  {journal} {Physical review letters}\ }\textbf {\bibinfo {volume} {107}},\ \bibinfo {pages} {240501} (\bibinfo {year} {2011})}\BibitemShut {NoStop}%
\bibitem [{\citenamefont {Devoret}\ and\ \citenamefont {Schoelkopf}(2013)}]{devoret2013superconducting}%
  \BibitemOpen
  \bibfield  {author} {\bibinfo {author} {\bibfnamefont {M.~H.}\ \bibnamefont {Devoret}}\ and\ \bibinfo {author} {\bibfnamefont {R.~J.}\ \bibnamefont {Schoelkopf}},\ }\href@noop {} {\bibfield  {journal} {\bibinfo  {journal} {Science}\ }\textbf {\bibinfo {volume} {339}},\ \bibinfo {pages} {1169} (\bibinfo {year} {2013})}\BibitemShut {NoStop}%
\bibitem [{\citenamefont {Sutherland}\ and\ \citenamefont {Srinivas}(2021)}]{sutherland2021universal}%
  \BibitemOpen
  \bibfield  {author} {\bibinfo {author} {\bibfnamefont {R.}~\bibnamefont {Sutherland}}\ and\ \bibinfo {author} {\bibfnamefont {R.}~\bibnamefont {Srinivas}},\ }\href@noop {} {\bibfield  {journal} {\bibinfo  {journal} {Physical Review A}\ }\textbf {\bibinfo {volume} {104}},\ \bibinfo {pages} {032609} (\bibinfo {year} {2021})}\BibitemShut {NoStop}%
\bibitem [{\citenamefont {Gan}\ \emph {et~al.}(2020)\citenamefont {Gan}, \citenamefont {Maslennikov}, \citenamefont {Tseng}, \citenamefont {Nguyen},\ and\ \citenamefont {Matsukevich}}]{gan2020hybrid}%
  \BibitemOpen
  \bibfield  {author} {\bibinfo {author} {\bibfnamefont {H.}~\bibnamefont {Gan}}, \bibinfo {author} {\bibfnamefont {G.}~\bibnamefont {Maslennikov}}, \bibinfo {author} {\bibfnamefont {K.-W.}\ \bibnamefont {Tseng}}, \bibinfo {author} {\bibfnamefont {C.}~\bibnamefont {Nguyen}},\ and\ \bibinfo {author} {\bibfnamefont {D.}~\bibnamefont {Matsukevich}},\ }\href@noop {} {\bibfield  {journal} {\bibinfo  {journal} {Physical review letters}\ }\textbf {\bibinfo {volume} {124}},\ \bibinfo {pages} {170502} (\bibinfo {year} {2020})}\BibitemShut {NoStop}%
\bibitem [{\citenamefont {Monroe}\ and\ \citenamefont {Kim}(2013)}]{monroe2013scaling}%
  \BibitemOpen
  \bibfield  {author} {\bibinfo {author} {\bibfnamefont {C.}~\bibnamefont {Monroe}}\ and\ \bibinfo {author} {\bibfnamefont {J.}~\bibnamefont {Kim}},\ }\href@noop {} {\bibfield  {journal} {\bibinfo  {journal} {Science}\ }\textbf {\bibinfo {volume} {339}},\ \bibinfo {pages} {1164} (\bibinfo {year} {2013})}\BibitemShut {NoStop}%
\bibitem [{\citenamefont {Wiersema}\ \emph {et~al.}(2020)\citenamefont {Wiersema}, \citenamefont {Zhou}, \citenamefont {de~Sereville}, \citenamefont {Carrasquilla}, \citenamefont {Kim},\ and\ \citenamefont {Yuen}}]{wiersema2020exploring}%
  \BibitemOpen
  \bibfield  {author} {\bibinfo {author} {\bibfnamefont {R.}~\bibnamefont {Wiersema}}, \bibinfo {author} {\bibfnamefont {C.}~\bibnamefont {Zhou}}, \bibinfo {author} {\bibfnamefont {Y.}~\bibnamefont {de~Sereville}}, \bibinfo {author} {\bibfnamefont {J.~F.}\ \bibnamefont {Carrasquilla}}, \bibinfo {author} {\bibfnamefont {Y.~B.}\ \bibnamefont {Kim}},\ and\ \bibinfo {author} {\bibfnamefont {H.}~\bibnamefont {Yuen}},\ }\href@noop {} {\bibfield  {journal} {\bibinfo  {journal} {PRX quantum}\ }\textbf {\bibinfo {volume} {1}},\ \bibinfo {pages} {020319} (\bibinfo {year} {2020})}\BibitemShut {NoStop}%
\bibitem [{\citenamefont {Johansson}\ \emph {et~al.}(2012)\citenamefont {Johansson}, \citenamefont {Nation},\ and\ \citenamefont {Nori}}]{johansson2012qutip}%
  \BibitemOpen
  \bibfield  {author} {\bibinfo {author} {\bibfnamefont {J.~R.}\ \bibnamefont {Johansson}}, \bibinfo {author} {\bibfnamefont {P.~D.}\ \bibnamefont {Nation}},\ and\ \bibinfo {author} {\bibfnamefont {F.}~\bibnamefont {Nori}},\ }\href@noop {} {\bibfield  {journal} {\bibinfo  {journal} {Computer physics communications}\ }\textbf {\bibinfo {volume} {183}},\ \bibinfo {pages} {1760} (\bibinfo {year} {2012})}\BibitemShut {NoStop}%
\bibitem [{\citenamefont {Hillery}\ \emph {et~al.}(1984)\citenamefont {Hillery}, \citenamefont {O'Connell}, \citenamefont {Scully},\ and\ \citenamefont {Wigner}}]{hillery1984distribution}%
  \BibitemOpen
  \bibfield  {author} {\bibinfo {author} {\bibfnamefont {M.}~\bibnamefont {Hillery}}, \bibinfo {author} {\bibfnamefont {R.~F.}\ \bibnamefont {O'Connell}}, \bibinfo {author} {\bibfnamefont {M.~O.}\ \bibnamefont {Scully}},\ and\ \bibinfo {author} {\bibfnamefont {E.~P.}\ \bibnamefont {Wigner}},\ }\href@noop {} {\bibfield  {journal} {\bibinfo  {journal} {Physics reports}\ }\textbf {\bibinfo {volume} {106}},\ \bibinfo {pages} {121} (\bibinfo {year} {1984})}\BibitemShut {NoStop}%
\bibitem [{\citenamefont {Breitenbach}\ \emph {et~al.}(1997)\citenamefont {Breitenbach}, \citenamefont {Schiller},\ and\ \citenamefont {Mlynek}}]{breitenbach1997measurement}%
  \BibitemOpen
  \bibfield  {author} {\bibinfo {author} {\bibfnamefont {G.}~\bibnamefont {Breitenbach}}, \bibinfo {author} {\bibfnamefont {S.}~\bibnamefont {Schiller}},\ and\ \bibinfo {author} {\bibfnamefont {J.}~\bibnamefont {Mlynek}},\ }\href@noop {} {\bibfield  {journal} {\bibinfo  {journal} {Nature}\ }\textbf {\bibinfo {volume} {387}},\ \bibinfo {pages} {471} (\bibinfo {year} {1997})}\BibitemShut {NoStop}%
\bibitem [{\citenamefont {Gietka}(2022)}]{gietka2022squeezing}%
  \BibitemOpen
  \bibfield  {author} {\bibinfo {author} {\bibfnamefont {K.}~\bibnamefont {Gietka}},\ }\href@noop {} {\bibfield  {journal} {\bibinfo  {journal} {Physical Review A}\ }\textbf {\bibinfo {volume} {105}},\ \bibinfo {pages} {042620} (\bibinfo {year} {2022})}\BibitemShut {NoStop}%
\bibitem [{\citenamefont {Wu}\ and\ \citenamefont {Hsieh}(2019)}]{wu2019variational}%
  \BibitemOpen
  \bibfield  {author} {\bibinfo {author} {\bibfnamefont {J.}~\bibnamefont {Wu}}\ and\ \bibinfo {author} {\bibfnamefont {T.~H.}\ \bibnamefont {Hsieh}},\ }\href@noop {} {\bibfield  {journal} {\bibinfo  {journal} {Physical review letters}\ }\textbf {\bibinfo {volume} {123}},\ \bibinfo {pages} {220502} (\bibinfo {year} {2019})}\BibitemShut {NoStop}%
\bibitem [{\citenamefont {Niu}\ \emph {et~al.}(2019)\citenamefont {Niu}, \citenamefont {Lu},\ and\ \citenamefont {Chuang}}]{niu2019optimizing}%
  \BibitemOpen
  \bibfield  {author} {\bibinfo {author} {\bibfnamefont {M.~Y.}\ \bibnamefont {Niu}}, \bibinfo {author} {\bibfnamefont {S.}~\bibnamefont {Lu}},\ and\ \bibinfo {author} {\bibfnamefont {I.~L.}\ \bibnamefont {Chuang}},\ }\href@noop {} {\bibfield  {journal} {\bibinfo  {journal} {arXiv preprint arXiv:1905.12134}\ } (\bibinfo {year} {2019})}\BibitemShut {NoStop}%
\bibitem [{\citenamefont {Ma}\ \emph {et~al.}(2011)\citenamefont {Ma}, \citenamefont {Wang}, \citenamefont {Sun},\ and\ \citenamefont {Nori}}]{ma2011quantum}%
  \BibitemOpen
  \bibfield  {author} {\bibinfo {author} {\bibfnamefont {J.}~\bibnamefont {Ma}}, \bibinfo {author} {\bibfnamefont {X.}~\bibnamefont {Wang}}, \bibinfo {author} {\bibfnamefont {C.-P.}\ \bibnamefont {Sun}},\ and\ \bibinfo {author} {\bibfnamefont {F.}~\bibnamefont {Nori}},\ }\href@noop {} {\bibfield  {journal} {\bibinfo  {journal} {Physics Reports}\ }\textbf {\bibinfo {volume} {509}},\ \bibinfo {pages} {89} (\bibinfo {year} {2011})}\BibitemShut {NoStop}%
\bibitem [{\citenamefont {Vidal}\ \emph {et~al.}(2003)\citenamefont {Vidal}, \citenamefont {Latorre}, \citenamefont {Rico},\ and\ \citenamefont {Kitaev}}]{Vidal_PRL_2003}%
  \BibitemOpen
  \bibfield  {author} {\bibinfo {author} {\bibfnamefont {G.}~\bibnamefont {Vidal}}, \bibinfo {author} {\bibfnamefont {J.~I.}\ \bibnamefont {Latorre}}, \bibinfo {author} {\bibfnamefont {E.}~\bibnamefont {Rico}},\ and\ \bibinfo {author} {\bibfnamefont {A.}~\bibnamefont {Kitaev}},\ }\href {https://doi.org/10.1103/PhysRevLett.90.227902} {\bibfield  {journal} {\bibinfo  {journal} {Phys. Rev. Lett.}\ }\textbf {\bibinfo {volume} {90}},\ \bibinfo {pages} {227902} (\bibinfo {year} {2003})}\BibitemShut {NoStop}%
\end{thebibliography}%

\end{document}